\documentclass[aps,nobibnotes,prl,superscriptaddress,amsmath,amssymb,amsfonts,twocolumn]{revtex4-1}
\usepackage{array,multirow,graphicx}
\usepackage{color}
\usepackage{ulem}
\usepackage[latin1]{inputenc}
\usepackage{caption}
\usepackage{scrextend}
\usepackage{hyperref}
\usepackage{float}
\usepackage{subcaption}
\usepackage{gensymb}

\def\bea{\begin{eqnarray}}
\def\eea{\end{eqnarray}}
\def\ben{\begin{equation}}
\def\een{\end{equation}}
\def\benu{\begin{enumerate}}
\def\enu{\end{enumerate}}
\def\bei{\begin{itemize}}
\def\eei{\end{itemize}}

\def\n{n}

\def\sss{\scriptscriptstyle\rm}

\def\1var{(\bx_1...\bx\N)}

\def\br{{\bf r}}

\def\bx{{\br t}}

\def\s{_{\sss S}}
\def\xc{_{\sss XC}}

\def\N{_{\sss N}}
\def\H{_{\sss H}}

\def\ext{_{\rm ext}}

\def\sph_int{ {\int d^3 r}}

\begin{document}

\normalem 

\title{Element specificity of transient extreme ultra-violet magnetic dichroism}
\author{J. K. Dewhurst}
\affiliation{Max-Planck-Institut f\"ur Mikrostrukturphysik, Weinberg 2, D-06120 Halle, Germany.}
\author{F. Willems}
\affiliation{Max-Born-Institut  f\"ur Nichtlineare Optik und Kurzzeitspektroskopie, Max-Born-Strasse 2A, 12489 Berlin, Germany.}
\author{P. Elliott}
\affiliation{Max-Born-Institut  f\"ur Nichtlineare Optik und Kurzzeitspektroskopie, Max-Born-Strasse 2A, 12489 Berlin, Germany.}
\author{Q. Z. Li}
\affiliation{Max-Born-Institut  f\"ur Nichtlineare Optik und Kurzzeitspektroskopie, Max-Born-Strasse 2A, 12489 Berlin, Germany.}
\author{C. von Korff Schmising}
\affiliation{Max-Born-Institut  f\"ur Nichtlineare Optik und Kurzzeitspektroskopie, Max-Born-Strasse 2A, 12489 Berlin, Germany.}
\author{C. Str\"uber}
\affiliation{Max-Born-Institut  f\"ur Nichtlineare Optik und Kurzzeitspektroskopie, Max-Born-Strasse 2A, 12489 Berlin, Germany.}
\author{D. W. Engel}
\affiliation{Max-Born-Institut  f\"ur Nichtlineare Optik und Kurzzeitspektroskopie, Max-Born-Strasse 2A, 12489 Berlin, Germany.}
\author{S. Eisebitt}
\affiliation{Max-Born-Institut  f\"ur Nichtlineare Optik und Kurzzeitspektroskopie, Max-Born-Strasse 2A, 12489 Berlin, Germany.}
\author{S. Sharma}
\affiliation{Max-Born-Institut  f\"ur Nichtlineare Optik und Kurzzeitspektroskopie, Max-Born-Strasse 2A, 12489 Berlin, Germany.}
\email{sharma@mbi-berlin.de}

\date{\today}

\begin{abstract}
In this work we combine theory and experiment to study transient magnetic circular dichroism (tr-MCD) in the extreme ultraviolet spectral range (XUV) in bulk Co and CoPt. We use the \emph{ab-initio} method of real-time time-dependent density functional theory (RT-TDDFT) to simulate the magnetization dynamics in the presence of ultrafast laser pulses. From this we demonstrate how tr-MCD may be calculated using an approximation to the excited-state linear-response. We apply this approximation to Co and CoPt and show computationally that element-specific dynamics of the local spin moments can be extracted from the tr-MCD in XUV energy range, as is commonly assumed. We then compare our theoretical prediction for the tr-MCD for CoPt with experimental measurement and find excellent agreement at many different frequencies including the $M_{2 3}$-edge of Co and $N_{6 7}$- and $O_{2 3}$- edges of Pt.
\end{abstract}

\maketitle

Ultrafast magnetization dynamics induced by femtosecond laser pulses is a rapidly developing research field. This is due to the vast increase in speed that such processes offer over traditional methods of magnetic manipulation. Due to short timescales and the push towards smaller lengthscales, the problem is intrinsically quantum-mechanical and thus challenging to understand and predict. In this regard, both theory and experiment must work together in order to comprehensively understand the problem.  

However in joint theory/experiment work there can be a  \emph{tower of babel} effect, where researchers misunderstand each other even when discussing the same physics. This is due to the fact that the quantities measured experimentally are often indirect measurements of those that are simulated. One such example is the field of femtomagnetism\cite{U09} where the magnetization dynamics is studied using magneto-optical probes such as magneto optical Kerr effect (MOKE)\cite{BMDB96}, second harmonic generation (mSHG)\cite{HMKB97}, and magnetic dichroism (MCD) \cite{SKPM07}. In these techniques it is assumed that the transient dynamics of the measured response is proportional to the magnetic moment. In particular, in MCD in the XUV energy range, it is assumed that the dynamics of specific peaks in the response function can be allocated to individual elements and thus observe element-specific magnetization dynamics. For this reason, MCD has become a powerful tool in probing spin dynamics. For example, the observation of a transient ferromagnetic state during all optical switching would not be possible without element specific probes\cite{RVSK11}.

In theoretical work on ultrafast spin dynamics, the fundamental quantity calculated is the spin magnetic moment itself. This is either found directly such as in Landau-Lifshitz-Gilbert (LLG) dynamics, or from the difference in the number of up/down spin electrons such as in the superdiffusive spin transport model\cite{BCO10} or the Boltzmann equation\cite{GFPZ12}, or via the expectation value of the Pauli matrices with spinor wavefunctions as is done in non-collinear real-time time-dependent density functional theory (TDDFT)\cite{KDES15}. Hence there is a disconnect between theory and experiment, which hinders direct comparison, and ultimately our understanding of ultrafast spin dynamics. Thus, a robust theoretical approach is required in order to bridge this divide. 

TDDFT is an \emph{ab-initio} simulation method for studying the charge and spin dynamics induced by laser pulses in realistic materials\cite{RG84,SDG14,EFB09}. In contrast to other theoretical approaches, the only input parameters are the atomic geometry and the laser pulse parameters. A Schr{\"o}dinger-like equation is then used to propagate 2-component Pauli spinors in time, from which the expectation value of observables, such as the magnetic moment, may be calculated. Real time (RT) TDDFT has proved successful in predicting and understanding ultrafast spin dynamics in bulk ferromagnetics\cite{KDES15,KEMS17}, Heusler compounds\cite{EMDS16}, Co/Cu interfaces\cite{CBEM19}, and Ni/Pt multilayers\cite{Siegrist2019,Dewhurst2018,DSS18}. Furthermore, linear response (LR) TDDFT was recently shown to successfully calculate the static MCD spectra for Fe, Co and Ni which requires a good description of the quantum mechanical electronic structure \cite{Willems2019}. Thus TDDFT is an ideal tool for connecting the worlds of experimental and theoretical spin dynamics. In this work we will use TDDFT to answer the question of whether element specific dynamics can be extracted from transient MCD spectra. In particular whether peaks can uniquely be assigned to individual elements, whether the dynamics of these peaks reproduce the local spin dynamics, and whether the dynamics of peaks at different frequencies agree. Furthermore we will compare the calculated MCD spectra directly with experiments.

Both of scientific and technological interest are multicomponent magnetic system such as CoPt as they show novel and intriguing functionalities when exposed to ultrashort optical excitation; both in the field of spintronics \cite{Kampfrath2013, Huisman2016} and for future all-optical data storage \cite{Lambert2014}. Element-specific and interface sensitive measurements in CoPt \cite{Willems2015} are therefore a prerequisite to understand the microscopic details of the interplay of the constituent elements, in particular as optically induced spin transfer (OISTR)\cite{EMDS16,DESG18} between Pt and Co
is expected. To study such rich and complex spin dynamics governed by a distinct magnetic response of Co and Pt atoms as well as between the Co atom in bulk Co and in CoPt, time resolved XUV MCD is an ideal experimental probe as it can access the relevant dichroic resonances in the 50-75 eV spectral range simultaneously in a single measurement. However, quantitative element-specificity is challenging to achieve as the relevant $M_{2 3}$-edge of Co and the $N_{6 7}$- and $O_{2 3}$- edges of Pt \cite{Shishidou1997a,Nakajima2002,Willems2017} are partly overlapped. Comparison of calculated tr-MCD spectra to both experiment and the calculated spin dynamics is therefore a requirement to assess the quality of calculated tr-MCD spectra as well as comment on the element specificity of tr-MCD technique itself.

\emph{Theoretical approach:} within LR-TDDFT, the linear response of the interacting system can be calculated\cite{SDG14} from that of the non-interacting Kohn-Sham (KS) system using the Dyson-like equation:
\ben
\label{dyson}
\chi(\omega) = \chi\s(\omega) + \chi_s(\omega) \left( v + f\xc \right)\chi(\omega),
\een
where $v$ is the bare Coulomb interaction and $f\xc$ is the exchange-correlation (XC) kernel.  The non-interacting response function, $\chi_s$, is given by:
\ben
\label{chis}
\chi\s(\br,\br',\omega) = \sum_{nmk} (f_{nk} -f_{mk})\frac{\phi^{*}_{nk}(\br)\phi^{*}_{mk}(\br')\phi_{nk}(\br')\phi_{mk}(\br)}{\omega - \epsilon_n + \epsilon_m + i\eta}, 
\een
where $f_{nk}$ is the occupation, $\epsilon_{nk}$ is the eigenvalue and $\phi_{nk}$ is the eigenvector of the $nk^{\rm th}$ KS orbital.
The dielectric response function can be calculated from this interacting response function, $\chi$, by using the relation:
\ben 
\label{diel}
\varepsilon^{-1}_{ij}(\omega) = \delta_{ij} + v\chi_{ij}(\omega).
\een
From $\varepsilon$ the magneto optical function\cite{OMSK92}, which is the experimental observable, can then be calculated using:
\ben
\Delta \delta (\omega) + i \Delta \beta (\omega) =  \frac{i\varepsilon_{xy}(\omega)}{2\sqrt{\varepsilon_{xx}(\omega)}}.
\een
This approach  has been very successful for calculating the static magneto-optical response function for Fe, Co and Ni\cite{Willems2019}.

Since by definition TDDFT must reproduce the exact change in the density of the system following a small perturbation, even when it is already driven by a pump laser, then the Dyson-like equation of Eq. \ref{dyson} must remain valid at any point in time. However, in order to find transient spectra one needs to modify the non-interacting response function, $\chi_s$. We approximate this using the transient occupations\cite{EMDS16} of the KS orbitals ($f_{nk}$), found by projecting the time-dependent KS orbitals into the ground-state orbitals:
\ben
\label{proj}
f_{nk}(t) = \sum_{m} f_{mk}(0) |\langle\psi_{mk}(t)|\phi_{nk}\rangle|^2,
\een
which can then be used with Eqs. \ref{dyson}, \ref{diel}, and \ref{chis}  to approximate the behavior of the dielectric as a function of time.
Formally this corresponds to linear response TDDFT of an excited-state wavefunction, but neglecting both memory and initial-state dependence of the kernel and approximating the KS response. In the following we will validate this approximation before comparing theoretical results to the experimental data. For the details of TDDFT and computational details we refer the reader to the supplementary  information.

\emph{Experimental technique:}
the static measurement of the magneto-optical function, $\Delta\beta$, was performed at the BESSY II synchrotron facility on the beamline UE112-PGM-1 in the XUV range from 45 eV to 75 eV.  Femtosecond pulses in the same spectral range were generated via high harmonic generation in neon, resulting in discrete harmonic emission peaks separated by $2\hbar\omega = 2 \times 1.55 $ eV. Here, photons are circularly polarized by a 4-mirror phase shifter\cite{Willems2015,vonKorffSchmising2017} and energetically dispersed after transmission through the magnetized sample. In both measurements we record the transmitted intensity ($I_\pm$) of circularly polarized photons through the magnetic sample for its two magnetization directions ($\pm$). We calculate the absorptive part of the magneto-optical function according to:
\ben
\label{expDeltabeta} 
\Delta\beta(\omega) = -\frac{c\tan\theta}{4d\omega}P\log(I_+/I_-)
\een
where $d$ is the thickness of the sample, $P$ the degree of circular polarization, $\theta$ the grazing angle of incidence and $I_{\pm}$ the absorption for left and right circularly polarized light. For details we refer to the supplementary information.

\begin{figure}[h]
{\includegraphics[width=0.5\textwidth]{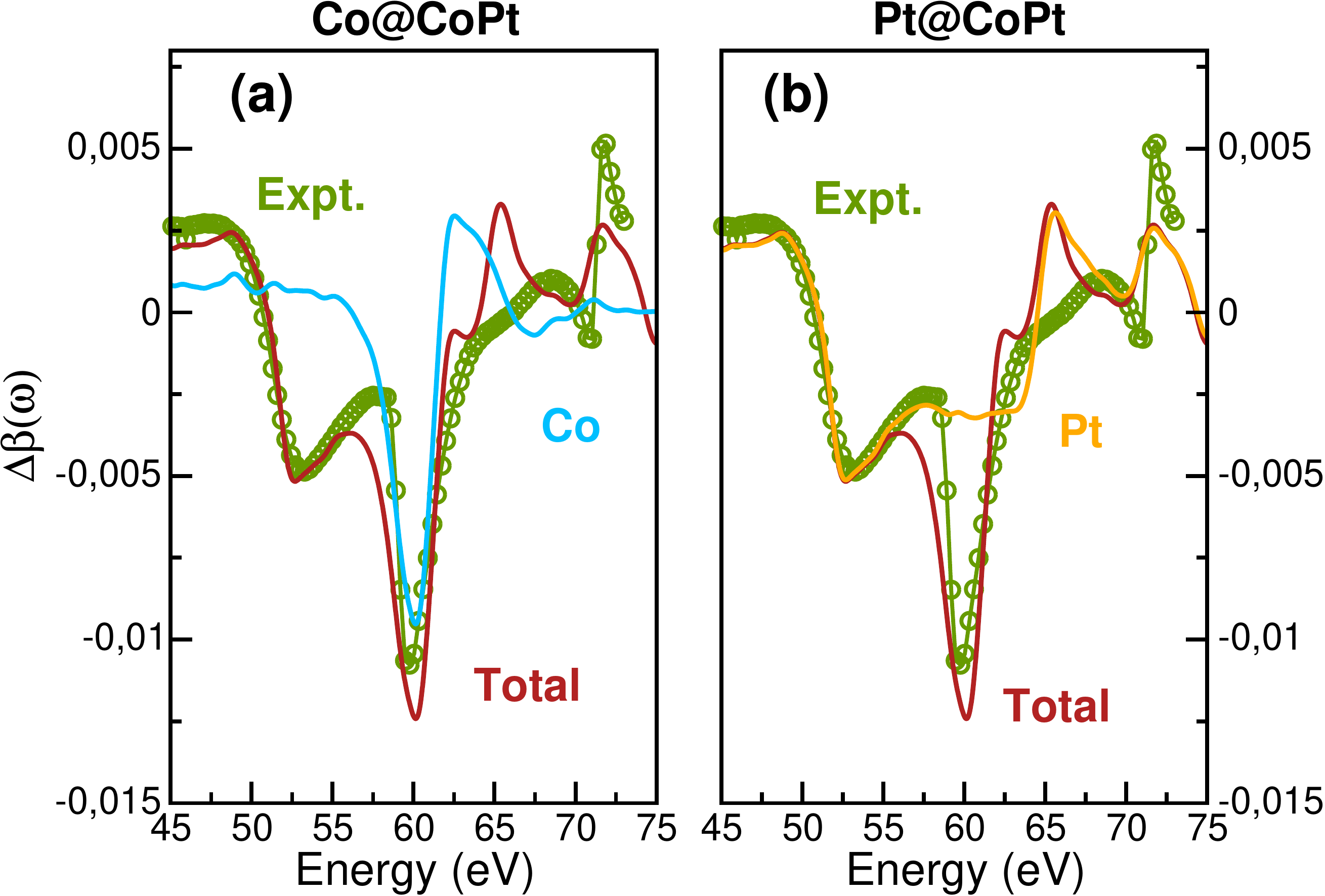}}
    \vspace{0.2cm}
    \caption{Ground-state $\Delta \beta$ for CoPt obtained experimentally (green dots) and by using TDDFT (red line). Theoretical data obtained by suppressing transitions from Pt $4f$ and $5p$ states is shown in (a) and by suppressing transitions from Co $3p$ states is shown in (b).}
\label{f:gs}    
\end{figure}
\emph{Results for the static case:} in Fig \ref{f:gs} we show the static MCD spectra for CoPt, both measured experimentally and computed via LR-TDDFT. As can be seen, the line-shape is well reproduced by LR-TDDFT-- the Co $M_{2 3}$-edge at 60 eV, Pt $O_{3}$-edge at 54 eV and Pt $N_{6 7}$-edge at 72 eV are very well reproduced by theoretical simulations. The Pt $O_{2}$-edge at 68 eV is red shifted by 3 eV in theory as compared to experiments. This extends the good agreement between experiment and theory found in Ref. \cite{Willems2019} for simple ferromagnets to the case of more complex materials. Furthermore, we can numerically decompose the total MCD signal into contributions for the Co and Pt individually. This is achieved by switching off the transitions from Pt $4f$ and $5p$ or Co $3p$ states for calculating the non-interacting response function of Eq. \ref{chis}. The pure Co contribution in CoPt, shown in Fig. \ref{f:gs}(a), has a pronounced peak at 60 eV, which is the $M_{2 3}$ edge corresponding to $3p\rightarrow 3d$ transitions. While the Pt contribution (Fig. \ref{f:gs}(b)) shows the $O_{2 3}$ peaks at 65 eV and 54 eV ($5p\rightarrow 5d$) with higher energy peaks coming from the $N_{6 7}$ edge around 72 eV ($4f\rightarrow 5d$).

However, it is important to note that there is cross-contamination between the Co and Pt edges. This can be clearly seen for the Co peak at 60 eV where Pt $O_{2 3}$-edges contributes significantly and for the Pt $O_{2 3}$-edge where transitions from Co states contribute significantly. This raises the question of whether the dynamics as traced by the intensity of the peak at 60 eV and 54 eV can corresponds to the dynamics of the Co and Pt local moment respectively, as is generally assumed to be the case\cite{Willems15,Hofherr18,Siegrist2019,Vorakiat2009,Mathias2012,Gang2018,Gunther2014}. 

\begin{figure}[h]
{\includegraphics[width=0.5\textwidth]{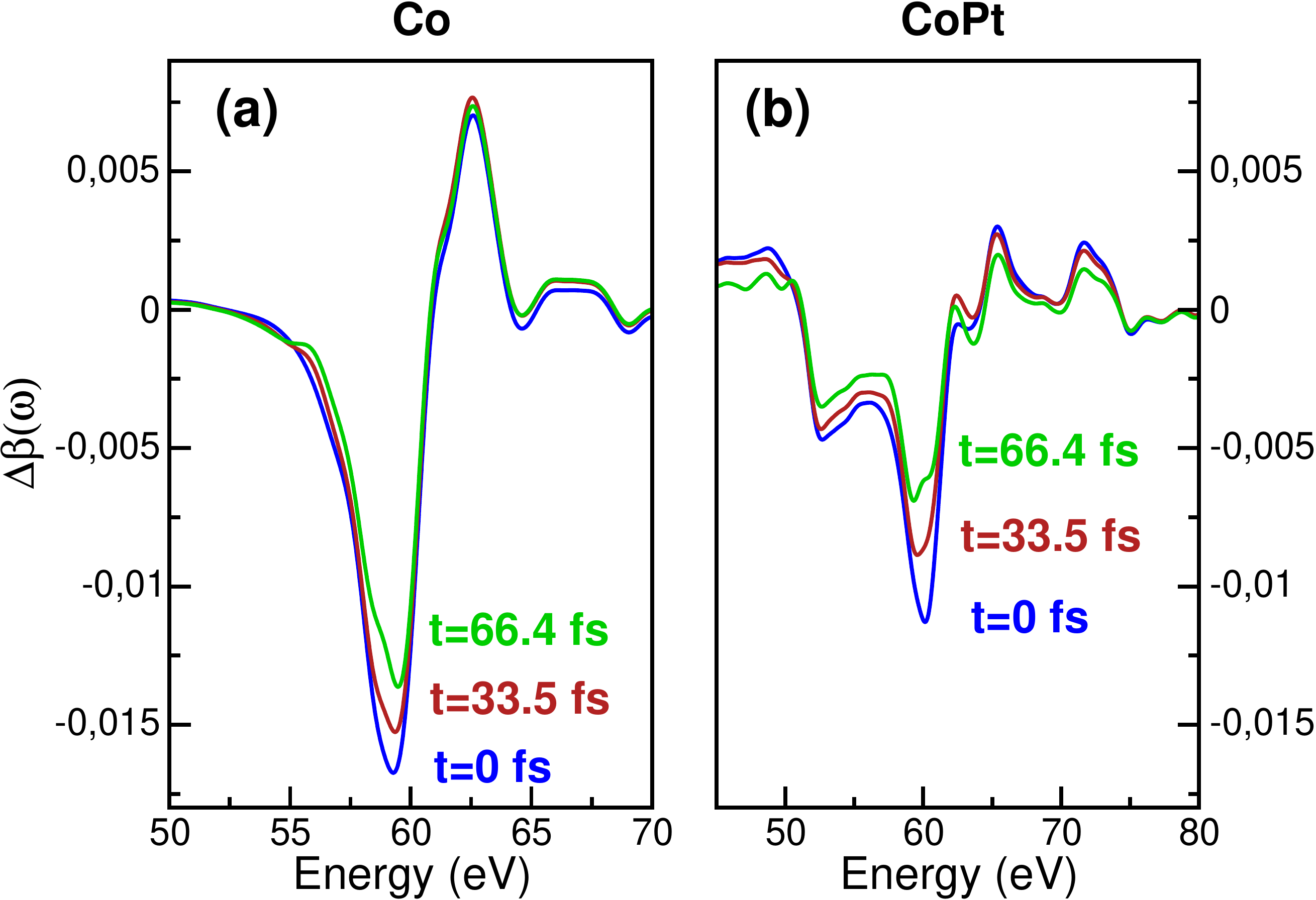}}
    \vspace{0.2cm}
    \caption{Transient $\Delta \beta$ for (a) bulk Co and (b) CoPt obtained using LR-TDDFT. The results for laser excited system are shown at various time steps. The pump laser used is linearly polarized red laser pulse with FWHM = 40 fs and incident fluence = 12 mJ/cm$^2$. }
\label{f:MCDt}
\end{figure}
\begin{figure}[h]
{\includegraphics[width=0.5\textwidth]{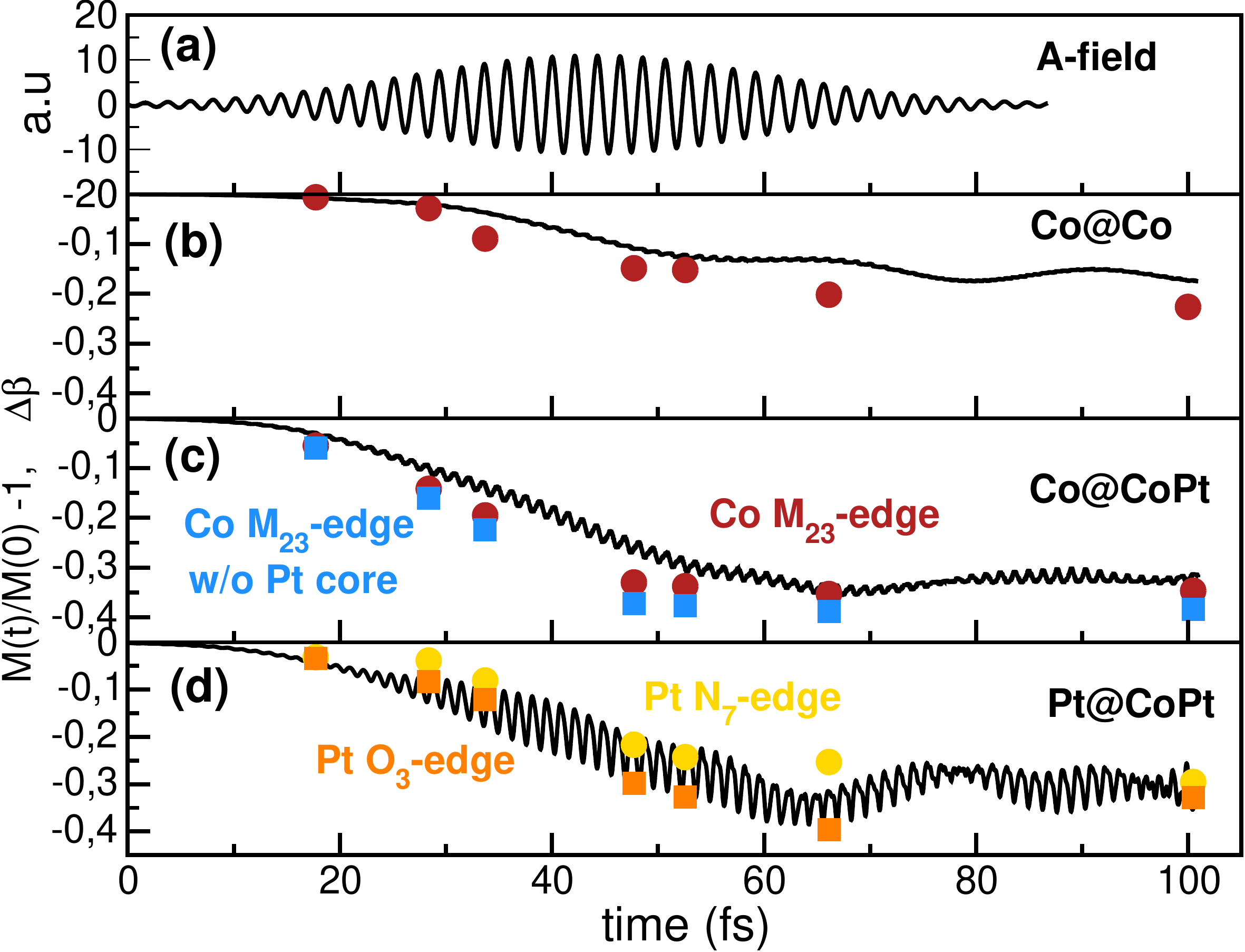}}
    \vspace{0.2cm}
    \caption{Magnetization dynamics of the laser excited system obtained using the dynamics of the peak heights in tr-MCD spectra (circles and squares) as compared to the one obtained using RT-TDDFT (full line). The vector potential of the pump laser pulse is shown in panel (a). Results for normalized magnetization are shown for (b) bulk Co, (c) Co in CoPt. Dynamics of Co peak at 60 eV obtained from the the total MCD (red circles) as well as the MCD spectra obtained by suppressing transitions from Pt $4f$ and $5p$ states (blue squares) is shown and (d) Pt in CoPt. Dynamics of Pt $O_{3}$-edge at 54 eV (orange squares) as well as $N_{7}$-edge (yellow circles) is shown. }
\label{f:magt}    
\end{figure}

\emph{Transient MCD:} having studied the MCD spectra for the static case, we now excite the system using an ultra-fast laser pulse and observe the dynamics of the magnetization and the MCD spectra. We choose a linearly polarized laser pulse with frequency $\omega=1.55$ eV, FWHM=40 fs, and a total incident fluence of 12 mJ/cm$^2$. The same pulse is used both experimentally and for theoretical simulations.
In Fig. \ref{f:MCDt} we show the evolution of the calculated MCD spectra at three different times during the simulation and compare it to the tr-MCD spectra of bulk Co. In all cases we see a decrease in the magnitude of the local moment (see Fig. \ref{f:magt}) which can be seen in the dynamics of the MCD signal. In the case of bulk Co, this demagnetization is due to the excitation of electrons from $3d$ to higher lying delocalized states as well as spin-orbit mediated spin-flips (as observed in Ref. \onlinecite{KDES15}). In the case of CoPt the dynamics is the results of two processes acting in tandem: (i) OISTR which causes the minority spins to transfer from the Pt to Co (as observed in Refs. \cite{Siegrist2019,Dewhurst2018}) and (ii) spin-orbit mediated spin-flips\cite{KDES15,KEMS17,ZH00,TP15,Kuiper14} which causes the Pt majority spins to flip into Pt as well as Co minority states leading to a further decrease in the local moment. 

We can extract the dynamics of the intensity of various MCD peaks calculated using LR-TDDFT  and compare this to the dynamics of the magnetization obtained using RT-TDDFT. In Fig. \ref{f:magt}, we first show the response of bulk Co where we find that the $M_{2 3}$-edge peak amplitude as a function of time reproduces the dynamics of the local spin moment. In the more complex CoPt system, we find that the Co $M_{2 3}$-edge peak follows the Co moment while both the $O_{3}$- and $N_{7}$-edges of Pt follow the Pt magnetization dynamics. This computationally validates (a) the approximation used for the non-interacting response function, indicating that the change in the occupation numbers is the most important contribution to the magnetization dynamics in early times and (b) the common interpretation of MCD data, which assigns the dynamics of specific peaks to the dynamics of specific elements.

For the ground-state MCD spectra, we saw that there was significant contribution from Pt $O_{2 3}$-edge to the Co peak at 60 eV. Despite this, we see that the dynamics of this peak could still follow the Co magnetization dynamics. To resolve this apparent contradiction, we plot in the Fig. \ref{f:magt}(b) the dynamics of the Co $M_{2 3}$-edge without the Pt core states (i.e. suppressing all transition from Pt $4f$ and $5p$ states in $\chi_s$). We find that since this cross-contamination does not significantly change over time, this does not affect the dynamics of the Co $M_{2 3}$-edge peak. Thus despite significant overlap of Co and Pt edges, element specific magnetization dynamics can be extracted from the MCD data. However, this is by no means guaranteed for all materials with overlapping edges. What one needs is a joint experimental and theoretical effort, where theory can help disentangle various experimental features by switching off element selective transitions.

\begin{figure}[h]
{\includegraphics[width=0.5\textwidth]{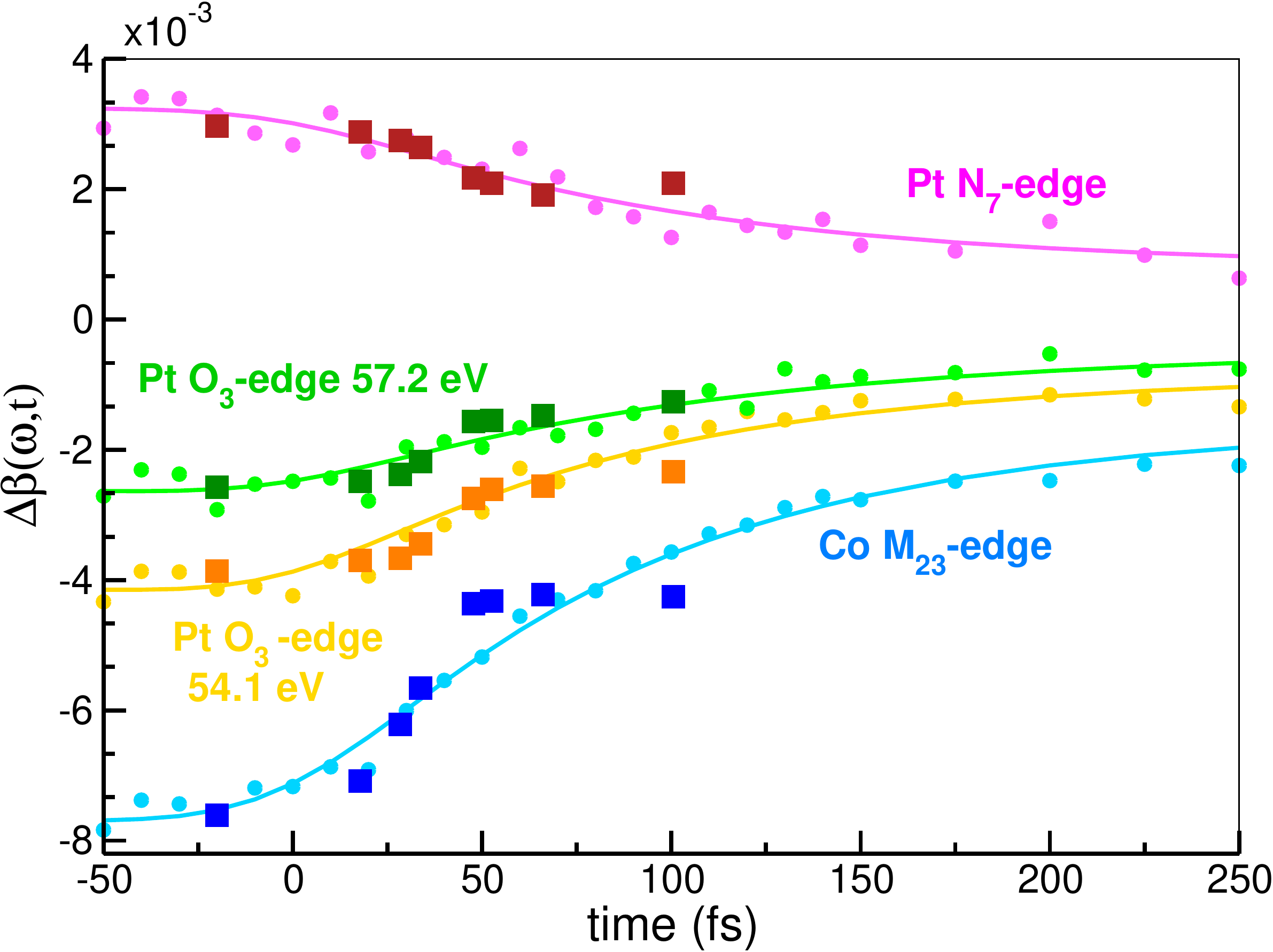}}
    \vspace{0.2cm}
    \caption{Dynamics of various features in theoretical tr-MCD (squares) compared to corresponding experimental data (circles with lines showing the exponential decay fitted to this data) for laser excited CoPt.} 
\label{f:cmpMCD}    
\end{figure}
\emph{Comparison of dynamics between theory and experiments:} having answered the question of whether the theoretical tr-MCD spectra may be used to extract element specific dynamics, we next compare our calculated tr-MCD spectra to the experimental data. This is shown in Fig. \ref{f:cmpMCD} at several XUV frequencies  corresponding to Co and Pt peaks. For the Co $M_{2 3}$-edge we see good agreement between theory and experiment. It is noteworthy that no scaling of the data was performed and we directly compare the experimental and theoretical $\Delta\beta(\omega)$. However, since there are slight deviations between the static theoretical and experimental peak intensities (see Fig. \ref{f:gs}) the data is rigidly shifted (in $y$-axis) to ensure that for t$<$0 the theoretical peak intensities match the experiment. 

For Pt, we examine three frequencies, the $O_{3}$-edge peak at 57.2 eV and 54.1 eV, and the $N_{7}$-edge peak at 72.1 eV which has opposite sign to the other two. In all cases, the TDDFT results show again an excellent agreement to the experimental data, even reproducing the difference in the magnitude of the response between the $O_{3}$- and $N_{7}$- edge peaks. 

Note, we performed our simulations for just the first ~100fs following the laser pulse. After this time lattice heating and electron-phonon interaction become dominant which causes further demagnetization,  however such processes are not included in our TDDFT simulations.

In conclusion, we used TDDFT to answer several questions regarding the interpretation of experimental transient MCD data. Firstly we demonstrated that there can be  cross-contamination of peaks in multi-element systems, casting doubt on whether element specific dynamics can be extracted from such data. However, based on our approximation to the transient MCD spectra using real-time TDDFT, we have removed this doubt. Specifically in CoPt, we show that the individual Co and Pt local spin moments in CoPt may be found by using the MCD peak dynamics at the respective $M_{2 3}$-, $N_{7}$- and $O_{3}$- edges. Finally we found excellent agreement in the transient behaviour as seen in our calculated tr-MCD spectra and the experimental data.

That MCD is element specific for complex magnets despite overlapping edges is by no means guaranteed, however a joint experimental and theoretical effort can help disentangle various experimental features. By developing a simple theoretical approach that allows for direct comparison between theory and experiment a more detailed understanding of the microscopic physical processes at play in ultrafast spin dynamics can now be achieved.  

\begin{acknowledgments}
SS, CvKS, PE, QZL and SE would like to thank DFG for funding through TRR227 projects A02 and A04.
\end{acknowledgments}

\newpage
\emph{Supplementary Information:}

\emph{Theoretical Details:}
TDDFT is an in-principle exact approach for calculating electron dynamics induced by external fields, such as laser pulses\cite{Runge1984,EFB09,SDG14}. It is the time-dependent extension of ground-state (GS) density functional theory (DFT) which provides an accurate and computationally efficient description of both the linear and non-linear response regimes. Both DFT and TDDFT map a system of interacting electrons to a  system of non-interacting electrons. This fictitious system is known as the Kohn-Sham (KS) system and defined such that it reproduces the same GS density (DFT) or density evolution (TDDFT) of the interacting system. It was proven, Hohenberg-Kohn theorem\cite{HK64} for DFT, Runge-Gross theorem\cite{Runge1984} for TDDFT, that knowledge of this density is sufficient in order to extract all observables of the system. To study spin dynamics, TDDFT is extended to reproduce the exact dynamics of the density and the magnetization density. For this case, the KS equation to be propagated is:
\bea\label{eq:fullksham}
\nonumber
&&i\frac{\partial \phi_j(\br,t)}{\partial t}=\left[
\frac{1}{2}\left(-i{\boldsymbol \nabla} +\frac{1}{c}{\bf A}\ext(t)\right)^2 \right. \\
&&+v\s(\br,t) + \frac{1}{2c} {\boldsymbol \sigma}\cdot{\bf B}_{s}(\br,t) +  \nonumber \\   && \left.
\frac{1}{4c^2} {\boldsymbol \sigma}\cdot ({\boldsymbol \nabla}v_{s}(\br,t) \times -i{\boldsymbol \nabla})\right]\phi_j(\br,t)
\eea
where $\phi_j(\br,t)$ are two-component Pauli spinors, ${\bf A}\ext(t)$ is the external 
laser field, written as a purely time-dependent vector potential, $\boldsymbol \sigma$ are the Pauli matrices, $v\s(\br,t)=v\ext(\br)+v\H(\br,t)+v\xc(\br,t)$ is the KS effective scalar potential, and ${\bf B}\s(\br,t)={\bf B}\ext(\br,t)+{\bf B}\xc(\br,t)$ is the KS effective magnetic field. These effective potentials ensure that the dynamics of the density, $\n(\br,t)$, and magnetization density, ${\bf m}(\br,t)$, is equal to that of the interacting system. The problem is defined by the external potentials, where the external scalar potential $v\ext(\br)$ includes the electron-nuclei interaction, while ${\bf B}\ext(\br,t)$ is any external magnetic field which interacts with the electronic spins via the Zeeman interaction.  The Hartree potential, $v\H(\br,t)$ is the classical electrostatic interaction. Finally, the XC potentials, the scalar $v\xc(\br,t)$, and the XC magnetic field, ${\bf B}\xc(\br,t)$, which require approximation. In the present work we use the adiabatic approximation, where these may be calculated using a DFT XC energy functional, $E\xc[\n,{\bf m}]$:
\bea
v\xc(\br,t) &=&  \left.\frac{\delta E\xc[\n,{\bf m}]}{\delta\n(\br)}\right|_{\n,{\bf m}=\n(\br,t),{\bf m}(\br,t)} \\
{\bf B}\xc(\br,t) &=& \left.\frac{\delta E\xc[\n,{\bf m}]}{\delta{\bf m}(\br)}\right|_{\n,{\bf m}=\n(\br,t),{\bf m}(\br,t)}.
\eea
The last term of  Eq.~(\ref{eq:fullksham}) is the spin-orbit coupling term.

\emph{Response function:} if the external perturbation is small, the linear response version of TDDFT can be used in the form of Eq.~(1) of the paper ($\chi(\omega) = \chi\s(\omega) + \chi_s(\omega) \left( v + f\xc \right)\chi(\omega)$). This is a matrix equation in reciprocal space of vectors {\bf G} i.e. $\chi$ which represents the response of the density to an external perturbation is a matrix. The reason for this is that an external perturbation, $e^{i({\bf G}+{\bf q}) \cdot {\bf r}}$ generates a response in the density of the form $e^{i({\bf G'}+{\bf q})\cdot{\bf r}}$. In order to solve this equation for $\chi$ one requires inversion of the matrix $1-(v+f_{\rm xc})\chi_s$ in {\bf G} space. This in turn allows for inclusion of the microscopic components known as the local field effects (LFE). These LFE are crucial for accurate description of the response function\cite{Willems2019}. In the present work these LFE are included and we found that we needed a matrix of $70 \times 70$ for convergence. It was also shown in Ref. \cite{Willems2019} that many-body correction to the KS band structure are important. In the present work we have included these corrections by first calculating a fully spin-polarized $GW$ spectral function to determine the correct position of the Co $3p$, Pt $4f$ and $5p$ states and then red shifting the KS states to these correct energies. 

\emph{Computational details:} all calculations are performed using the highly accurate full potential linearized augmented-plane-wave method\cite{singh}, as implemented in the ELK\cite{elk} code. A smearing width of 0.027eV was used for ground-state and RT-TDDFT calculations and a smearing of 1 eV was used for linear response calculations. All states greater than 95~eV below Fermi-level were treated as Dirac spinors, i.e. obtained by solving the Dirac equation. All the other states are treated as Pauli spinors obtained by solving the Schr\"odinger equation including spin-orbit and other relativistic corrections (e.g mass correction and Darwin terms). All states up to 90~eV above the Fermi level were included in the calculations.
A bulk FCC Co unit cell with lattice parameter of $3.21 \AA$ was used. An ordered 50-50 alloy for CoPt was simulated using a L1$_0$ unit cell with lattice parameter of $a=7.2\AA, c=7.0\AA$. The Brillouin zone was sampled with a $8\times 8\times 8$ mesh for CoPt and $10\times 10\times 10$ for bulk Co. The XC energy functional was LDA\cite{lda}, which was extended to treat non-collinear systems according to the method of K{\"u}bler\cite{KHSW88}. For time propagation the algorithm detailed in Ref. \onlinecite{Dewhurst2016} was used with a time-step of $2.42$ atto-seconds. 

The final magnetization value for each atom is converged with these parameters. We obtained a magnetic moment of $1.66 \mu_{\rm B}$ per atom for bulk Co. In the case of CoPt a moment of $1.877 \mu_{\rm B}$ on Co atoms and an induced moment of $0.38 \mu_{\rm B}$ on Pt atom was obtained.  
During the time propagation we see that there are small oscillations (with large period) around the final value of the moment (see Fig. 3 of the manuscript). These oscillations are numerical and get damped as you increase the number of {\bf k}-points. In contrast to this, the rapid oscillations seen in Figs. 3(b)-(d) of the manuscript are due to the electrons moving back and forth with the frequency of
the electric field (as well as higher harmonics). The local moments are extracted by integration of the magnetization within a sphere around the each atom and this leads to a doubling of the frequency of any oscillation and hence the frequency in Figs. 3(b)-(d) of the manuscript is twice that of the pump-pulse frequency in Fig. 3(a).

At this point it is important to note the difference between the results for bulk Co in the present case and in the previous work in Ref. \onlinecite{Willems2019}-- there are three differences: (i) in the present work an average width of 1 eV for $3p_{1/2}$ and $3p_{3/2}$ states is considered, while in the previous work a state dependent width was used. The net result of this is that the intensity of the post $M_{2 3}$-edge peak is large, (ii) in the present work states up to 90 eV above the Fermi level are used, while in the previous work states up to 300 eV above the Fermi level were used. The net result of this that the pre-edge features are suppressed and (iii) a {\bf k}-point mesh of $10\times 10\times 10$ is used in the present work, which is half of what was used before. The reason for all these is the computational demand: in the present case we need to perform a full time propagation which is computationally very demanding and memory intensive.

\emph{Experimental Details:}
\begin{figure}[h]
{\includegraphics[width=0.5\textwidth]{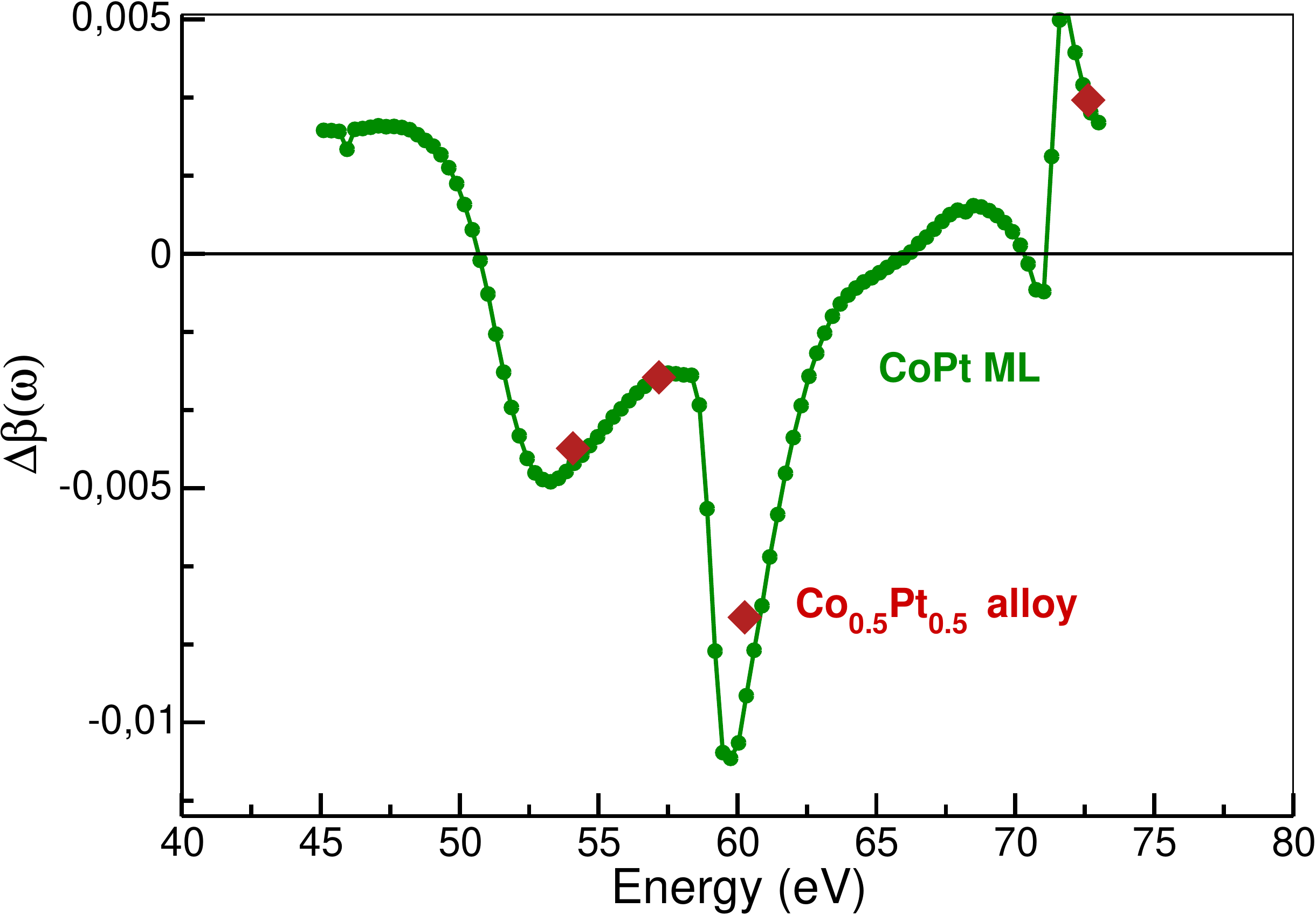}}
    \vspace{0.2cm}
    \caption{$\Delta \beta(\omega)$ obtained experimentally at the synchrotron facility BESSY II for the Co/Pt multilayer (green dots) and measured with a high harmonic source at selected photon energies of the Co$_{0.5}$Pt$_0.5$ alloy. The good agreement supports our approximation of the effective thickness of magnetized Pt in the Co/Pt multi-layer.}
\label{f:gs}
\end{figure}

The Co/Pt multilayer, [Co(0.4 nm)/Pt(0.7 nm)]$_{30}$, and the Co$_{0.5}$Pt$_{0.5}$(15 nm) alloy samples were grown by magnetron sputtering on free-standing Si$_3$N$_4$ (20 nm thickness) and aluminum (200 nm thickness) membranes, respectively.   The Co/Pt multi-layer has an out-of-plane anisotropy with a coercive field of 250 mT, while the CoPt alloy is magnetized in-plane and has a smaller coercive field of $<$ 10 mT.

The static measurement of the absorptive part of the magneto-optical function, $\Delta\beta$, was measured for the multi-layer [Co(0.4 nm)/Pt(0.7 nm)]$_{30}$ at the undulator beamline UE112-PGM1 of BESSY II, HZB, with variable polarization and an energy resolution $E/\Delta E$ = 30000 in the spectral range from 45 eV to 75 eV (green dots in Fig. \ref{f:gs}). Here, we confirmed the degree of circular polarization of $P_{\rm{circ}} = 0.99$ in an independent measurement. To calculate $\Delta\beta(\omega)$ we estimate that the induced magnetic moment of the Pt atoms extends 0.2 nm from each Co/Pt interface \cite{Suzuki2005}, such that we can approximate an identical effective thickness of Co and Pt of $d = 30 \cdot 0.4$ nm. Importantly, this implies that the \emph{static} $\Delta\beta$ is comparable to a Co$_{0.5}$Pt$_{0.5}$ alloy.
Time and frequency resolved $\Delta\beta(\omega, t)$ as shown in Fig. 4 was measured in an optical pump (fluence 12 mJ/cm$^2$, center photon energy $\hbar \omega = 1.55 eV$, pulse duration $\Delta\tau = 50$ fs), resonant XUV probe experiment in an collinear geometry. Femtosecond pulses in the extreme ultraviolet spectral range were generated via high harmonic generation by focusing laser pulses (center photon energy $\hbar \omega = 1.55 eV$, pulse duration $\Delta\tau = 30$ fs, pulse energy $E = 2.5$ mJ, repetition rate $\nu = 3 $ kHz) into a neon gas cell. The discrete, odd harmonics with a spectral width of approximately 200 meV are transmitted through the sample, energetically dispersed by a reflection grating and detected  by an XUV-sensitive charge-coupled device camera. The sample is mounted under a grazing angle of $\theta = 50$ degree for a finite projection of the \textbf{k} vector of the circularly polarized XUV pulses and the magnetization of the sample. We integrate the transmitted intensity, $I_{\rm{\pm}}$, of one harmonic and calculate $\Delta\beta$  according to Eqn. (8), taking into account the finite and energy dependent degree of polarization \cite{Willems2015,vonKorffSchmising2017} and the angle of incidence, but ignoring smaller corrections due to refraction at the vacuum/sample interface \cite{Willems2019}.
$\Delta\beta(t<0)$ of the CoPt alloy as determined in the time resolved high harmonic experiment are also shown in Fig. 1 (red diamonds) and show an excellent agreement with the static measurements, corroborating our approximation of the effective Pt thickness within the Co/Pt multilayer.


\begin{thebibliography}{45}%
\makeatletter
\providecommand \@ifxundefined [1]{%
 \@ifx{#1\undefined}
}%
\providecommand \@ifnum [1]{%
 \ifnum #1\expandafter \@firstoftwo
 \else \expandafter \@secondoftwo
 \fi
}%
\providecommand \@ifx [1]{%
 \ifx #1\expandafter \@firstoftwo
 \else \expandafter \@secondoftwo
 \fi
}%
\providecommand \natexlab [1]{#1}%
\providecommand \enquote  [1]{``#1''}%
\providecommand \bibnamefont  [1]{#1}%
\providecommand \bibfnamefont [1]{#1}%
\providecommand \citenamefont [1]{#1}%
\providecommand \href@noop [0]{\@secondoftwo}%
\providecommand \href [0]{\begingroup \@sanitize@url \@href}%
\providecommand \@href[1]{\@@startlink{#1}\@@href}%
\providecommand \@@href[1]{\endgroup#1\@@endlink}%
\providecommand \@sanitize@url [0]{\catcode `\\12\catcode `\$12\catcode
  `\&12\catcode `\#12\catcode `\^12\catcode `\_12\catcode `\%12\relax}%
\providecommand \@@startlink[1]{}%
\providecommand \@@endlink[0]{}%
\providecommand \url  [0]{\begingroup\@sanitize@url \@url }%
\providecommand \@url [1]{\endgroup\@href {#1}{\urlprefix }}%
\providecommand \urlprefix  [0]{URL }%
\providecommand \Eprint [0]{\href }%
\providecommand \doibase [0]{http://dx.doi.org/}%
\providecommand \selectlanguage [0]{\@gobble}%
\providecommand \bibinfo  [0]{\@secondoftwo}%
\providecommand \bibfield  [0]{\@secondoftwo}%
\providecommand \translation [1]{[#1]}%
\providecommand \BibitemOpen [0]{}%
\providecommand \bibitemStop [0]{}%
\providecommand \bibitemNoStop [0]{.\EOS\space}%
\providecommand \EOS [0]{\spacefactor3000\relax}%
\providecommand \BibitemShut  [1]{\csname bibitem#1\endcsname}%
\let\auto@bib@innerbib\@empty
\bibitem [{\citenamefont {Bovensiepen}(2009)}]{U09}%
  \BibitemOpen
  \bibfield  {author} {\bibinfo {author} {\bibfnamefont {U.}~\bibnamefont
  {Bovensiepen}},\ }\href@noop {} {\bibfield  {journal} {\bibinfo  {journal}
  {Nat. Phys.}\ }\textbf {\bibinfo {volume} {5}},\ \bibinfo {pages} {461}
  (\bibinfo {year} {2009})}\BibitemShut {NoStop}%
\bibitem [{\citenamefont {Beaurepaire}\ \emph {et~al.}(1996)\citenamefont
  {Beaurepaire}, \citenamefont {Merle}, \citenamefont {Daunois},\ and\
  \citenamefont {Bigot}}]{BMDB96}%
  \BibitemOpen
  \bibfield  {author} {\bibinfo {author} {\bibfnamefont {E.}~\bibnamefont
  {Beaurepaire}}, \bibinfo {author} {\bibfnamefont {J.}~\bibnamefont {Merle}},
  \bibinfo {author} {\bibfnamefont {A.}~\bibnamefont {Daunois}}, \ and\
  \bibinfo {author} {\bibfnamefont {J.}~\bibnamefont {Bigot}},\ }\href
  {\doibase 10.1103/PhysRevLett.76.4250} {\bibfield  {journal} {\bibinfo
  {journal} {Phys. Rev. Lett.}\ }\textbf {\bibinfo {volume} {76}},\ \bibinfo
  {pages} {4250} (\bibinfo {year} {1996})}\BibitemShut {NoStop}%
\bibitem [{\citenamefont {Hohlfeld}\ \emph {et~al.}(1997)\citenamefont
  {Hohlfeld}, \citenamefont {Matthias}, \citenamefont {Knorren},\ and\
  \citenamefont {Bennemann}}]{HMKB97}%
  \BibitemOpen
  \bibfield  {author} {\bibinfo {author} {\bibfnamefont {J.}~\bibnamefont
  {Hohlfeld}}, \bibinfo {author} {\bibfnamefont {E.}~\bibnamefont {Matthias}},
  \bibinfo {author} {\bibfnamefont {R.}~\bibnamefont {Knorren}}, \ and\
  \bibinfo {author} {\bibfnamefont {K.}~\bibnamefont {Bennemann}},\ }\href
  {\doibase 10.1103/PhysRevLett.78.4861} {\bibfield  {journal} {\bibinfo
  {journal} {Phys. Rev. Lett.}\ }\textbf {\bibinfo {volume} {78}},\ \bibinfo
  {pages} {4861} (\bibinfo {year} {1997})}\BibitemShut {NoStop}%
\bibitem [{\citenamefont {Stamm}\ \emph {et~al.}(2007)\citenamefont {Stamm},
  \citenamefont {Kachel}, \citenamefont {Pontius}, \citenamefont {Mitzner},
  \citenamefont {Quast}, \citenamefont {Holldack}, \citenamefont {Khan},
  \citenamefont {Lupulescu}, \citenamefont {Aziz}, \citenamefont {Wietstruk},
  \citenamefont {Durr},\ and\ \citenamefont {Eberhardt}}]{SKPM07}%
  \BibitemOpen
  \bibfield  {author} {\bibinfo {author} {\bibfnamefont {C.}~\bibnamefont
  {Stamm}}, \bibinfo {author} {\bibfnamefont {T.}~\bibnamefont {Kachel}},
  \bibinfo {author} {\bibfnamefont {N.}~\bibnamefont {Pontius}}, \bibinfo
  {author} {\bibfnamefont {R.}~\bibnamefont {Mitzner}}, \bibinfo {author}
  {\bibfnamefont {T.}~\bibnamefont {Quast}}, \bibinfo {author} {\bibfnamefont
  {K.}~\bibnamefont {Holldack}}, \bibinfo {author} {\bibfnamefont
  {S.}~\bibnamefont {Khan}}, \bibinfo {author} {\bibfnamefont {C.}~\bibnamefont
  {Lupulescu}}, \bibinfo {author} {\bibfnamefont {E.~F.}\ \bibnamefont {Aziz}},
  \bibinfo {author} {\bibfnamefont {M.}~\bibnamefont {Wietstruk}}, \bibinfo
  {author} {\bibfnamefont {H.~A.}\ \bibnamefont {Durr}}, \ and\ \bibinfo
  {author} {\bibfnamefont {W.}~\bibnamefont {Eberhardt}},\ }\href@noop {}
  {\bibfield  {journal} {\bibinfo  {journal} {Nat. Mater.}\ }\textbf {\bibinfo
  {volume} {6}},\ \bibinfo {pages} {740} (\bibinfo {year} {2007})}\BibitemShut
  {NoStop}%
\bibitem [{\citenamefont {Radu}\ \emph {et~al.}(2011)\citenamefont {Radu},
  \citenamefont {Vahaplar}, \citenamefont {Stamm}, \citenamefont {Kachel},
  \citenamefont {Pontius}, \citenamefont {Duerr}, \citenamefont {Ostler},
  \citenamefont {Barker}, \citenamefont {Evans}, \citenamefont {Chantrell},
  \citenamefont {Tsukamoto}, \citenamefont {Itoh}, \citenamefont {Kirilyuk},
  \citenamefont {Rasing},\ and\ \citenamefont {Kimel}}]{RVSK11}%
  \BibitemOpen
  \bibfield  {author} {\bibinfo {author} {\bibfnamefont {I.}~\bibnamefont
  {Radu}}, \bibinfo {author} {\bibfnamefont {K.}~\bibnamefont {Vahaplar}},
  \bibinfo {author} {\bibfnamefont {C.}~\bibnamefont {Stamm}}, \bibinfo
  {author} {\bibfnamefont {T.}~\bibnamefont {Kachel}}, \bibinfo {author}
  {\bibfnamefont {N.}~\bibnamefont {Pontius}}, \bibinfo {author} {\bibfnamefont
  {H.~A.}\ \bibnamefont {Duerr}}, \bibinfo {author} {\bibfnamefont {T.~A.}\
  \bibnamefont {Ostler}}, \bibinfo {author} {\bibfnamefont {J.}~\bibnamefont
  {Barker}}, \bibinfo {author} {\bibfnamefont {R.~F.~L.}\ \bibnamefont
  {Evans}}, \bibinfo {author} {\bibfnamefont {R.~W.}\ \bibnamefont
  {Chantrell}}, \bibinfo {author} {\bibfnamefont {A.}~\bibnamefont
  {Tsukamoto}}, \bibinfo {author} {\bibfnamefont {A.}~\bibnamefont {Itoh}},
  \bibinfo {author} {\bibfnamefont {A.}~\bibnamefont {Kirilyuk}}, \bibinfo
  {author} {\bibfnamefont {T.}~\bibnamefont {Rasing}}, \ and\ \bibinfo {author}
  {\bibfnamefont {A.~V.}\ \bibnamefont {Kimel}},\ }\href {\doibase
  10.1038/nature09901} {\bibfield  {journal} {\bibinfo  {journal} {Nature}\
  }\textbf {\bibinfo {volume} {472}},\ \bibinfo {pages} {205} (\bibinfo {year}
  {2011})}\BibitemShut {NoStop}%
\bibitem [{\citenamefont {Battiato}\ \emph {et~al.}(2010)\citenamefont
  {Battiato}, \citenamefont {Carva},\ and\ \citenamefont {Oppeneer}}]{BCO10}%
  \BibitemOpen
  \bibfield  {author} {\bibinfo {author} {\bibfnamefont {M.}~\bibnamefont
  {Battiato}}, \bibinfo {author} {\bibfnamefont {K.}~\bibnamefont {Carva}}, \
  and\ \bibinfo {author} {\bibfnamefont {P.~M.}\ \bibnamefont {Oppeneer}},\
  }\href@noop {} {\bibfield  {journal} {\bibinfo  {journal} {Phys. Rev. Lett.}\
  }\textbf {\bibinfo {volume} {105}},\ \bibinfo {pages} {027203} (\bibinfo
  {year} {2010})}\BibitemShut {NoStop}%
\bibitem [{\citenamefont {Gradhand}\ \emph {et~al.}(2012)\citenamefont
  {Gradhand}, \citenamefont {Fedorov}, \citenamefont {Pientka}, \citenamefont
  {Zahn}, \citenamefont {Mertig},\ and\ \citenamefont {Gy{\"o}rffy}}]{GFPZ12}%
  \BibitemOpen
  \bibfield  {author} {\bibinfo {author} {\bibfnamefont {M.}~\bibnamefont
  {Gradhand}}, \bibinfo {author} {\bibfnamefont {D.~V.}\ \bibnamefont
  {Fedorov}}, \bibinfo {author} {\bibfnamefont {F.}~\bibnamefont {Pientka}},
  \bibinfo {author} {\bibfnamefont {P.}~\bibnamefont {Zahn}}, \bibinfo {author}
  {\bibfnamefont {I.}~\bibnamefont {Mertig}}, \ and\ \bibinfo {author}
  {\bibfnamefont {B.~L.}\ \bibnamefont {Gy{\"o}rffy}},\ }\href {\doibase
  10.1088/0953-8984/24/21/213202} {\bibfield  {journal} {\bibinfo  {journal}
  {Journal of Physics: Condensed Matter}\ }\textbf {\bibinfo {volume} {24}},\
  \bibinfo {pages} {213202} (\bibinfo {year} {2012})}\BibitemShut {NoStop}%
\bibitem [{\citenamefont {Krieger}\ \emph {et~al.}(2015)\citenamefont
  {Krieger}, \citenamefont {Dewhurst}, \citenamefont {Elliott}, \citenamefont
  {Sharma},\ and\ \citenamefont {Gross}}]{KDES15}%
  \BibitemOpen
  \bibfield  {author} {\bibinfo {author} {\bibfnamefont {K.}~\bibnamefont
  {Krieger}}, \bibinfo {author} {\bibfnamefont {J.~K.}\ \bibnamefont
  {Dewhurst}}, \bibinfo {author} {\bibfnamefont {P.}~\bibnamefont {Elliott}},
  \bibinfo {author} {\bibfnamefont {S.}~\bibnamefont {Sharma}}, \ and\ \bibinfo
  {author} {\bibfnamefont {E.~K.~U.}\ \bibnamefont {Gross}},\ }\href {\doibase
  10.1021/acs.jctc.5b00621} {\bibfield  {journal} {\bibinfo  {journal} {J.
  Chem. Theory Comput.}\ }\textbf {\bibinfo {volume} {11}},\ \bibinfo {pages}
  {4870} (\bibinfo {year} {2015})}\BibitemShut {NoStop}%
\bibitem [{\citenamefont {Runge}\ and\ \citenamefont
  {Gross}(1984{\natexlab{a}})}]{RG84}%
  \BibitemOpen
  \bibfield  {author} {\bibinfo {author} {\bibfnamefont {E.}~\bibnamefont
  {Runge}}\ and\ \bibinfo {author} {\bibfnamefont {E.}~\bibnamefont {Gross}},\
  }\href@noop {} {\bibfield  {journal} {\bibinfo  {journal} {Phys. Rev. Lett.}\
  }\textbf {\bibinfo {volume} {52}},\ \bibinfo {pages} {997} (\bibinfo {year}
  {1984}{\natexlab{a}})}\BibitemShut {NoStop}%
\bibitem [{\citenamefont {Sharma}\ \emph {et~al.}(2014)\citenamefont {Sharma},
  \citenamefont {Dewhurst},\ and\ \citenamefont {Gross}}]{SDG14}%
  \BibitemOpen
  \bibfield  {author} {\bibinfo {author} {\bibfnamefont {S.}~\bibnamefont
  {Sharma}}, \bibinfo {author} {\bibfnamefont {J.~K.}\ \bibnamefont
  {Dewhurst}}, \ and\ \bibinfo {author} {\bibfnamefont {E.~K.~U.}\ \bibnamefont
  {Gross}},\ }\enquote {\bibinfo {title} {Optical response of extended systems
  using time-dependent density functional theory},}\ in\ \href {\doibase
  10.1007/128_2014_529} {\emph {\bibinfo {booktitle} {First Principles
  Approaches to Spectroscopic Properties of Complex Materials}}},\ \bibinfo
  {editor} {edited by\ \bibinfo {editor} {\bibfnamefont {C.}~\bibnamefont
  {Di~Valentin}}, \bibinfo {editor} {\bibfnamefont {S.}~\bibnamefont {Botti}},
  \ and\ \bibinfo {editor} {\bibfnamefont {M.}~\bibnamefont {Cococcioni}}}\
  (\bibinfo  {publisher} {Springer Berlin Heidelberg},\ \bibinfo {address}
  {Berlin, Heidelberg},\ \bibinfo {year} {2014})\ pp.\ \bibinfo {pages}
  {235--257}\BibitemShut {NoStop}%
\bibitem [{\citenamefont {Elliott}\ \emph {et~al.}(2009)\citenamefont
  {Elliott}, \citenamefont {Furche},\ and\ \citenamefont {Burke}}]{EFB09}%
  \BibitemOpen
  \bibfield  {author} {\bibinfo {author} {\bibfnamefont {P.}~\bibnamefont
  {Elliott}}, \bibinfo {author} {\bibfnamefont {F.}~\bibnamefont {Furche}}, \
  and\ \bibinfo {author} {\bibfnamefont {K.}~\bibnamefont {Burke}},\ }in\
  \href@noop {} {\emph {\bibinfo {booktitle} {Reviews in Computational
  Chemistry}}},\ Vol.~\bibinfo {volume} {26},\ \bibinfo {editor} {edited by\
  \bibinfo {editor} {\bibfnamefont {K.}~\bibnamefont {Lipkowitz}}\ and\
  \bibinfo {editor} {\bibfnamefont {T.}~\bibnamefont {Cundari}}}\ (\bibinfo
  {publisher} {Wiley, Hoboken, NJ},\ \bibinfo {year} {2009})\ pp.\ \bibinfo
  {pages} {91--165}\BibitemShut {NoStop}%
\bibitem [{\citenamefont {Krieger}\ \emph {et~al.}(2017)\citenamefont
  {Krieger}, \citenamefont {Elliott}, \citenamefont {M{\"u}ller}, \citenamefont
  {Singh}, \citenamefont {Dewhurst}, \citenamefont {Gross},\ and\ \citenamefont
  {Sharma}}]{KEMS17}%
  \BibitemOpen
  \bibfield  {author} {\bibinfo {author} {\bibfnamefont {K.}~\bibnamefont
  {Krieger}}, \bibinfo {author} {\bibfnamefont {P.}~\bibnamefont {Elliott}},
  \bibinfo {author} {\bibfnamefont {T.}~\bibnamefont {M{\"u}ller}}, \bibinfo
  {author} {\bibfnamefont {N.}~\bibnamefont {Singh}}, \bibinfo {author}
  {\bibfnamefont {J.~K.}\ \bibnamefont {Dewhurst}}, \bibinfo {author}
  {\bibfnamefont {E.~K.~U.}\ \bibnamefont {Gross}}, \ and\ \bibinfo {author}
  {\bibfnamefont {S.}~\bibnamefont {Sharma}},\ }\href
  {http://stacks.iop.org/0953-8984/29/i=22/a=224001} {\bibfield  {journal}
  {\bibinfo  {journal} {Journal of Physics: Condensed Matter}\ }\textbf
  {\bibinfo {volume} {29}},\ \bibinfo {pages} {224001} (\bibinfo {year}
  {2017})}\BibitemShut {NoStop}%
\bibitem [{\citenamefont {Elliott}\ \emph {et~al.}(2016)\citenamefont
  {Elliott}, \citenamefont {Mueller}, \citenamefont {Dewhurst}, \citenamefont
  {Sharma},\ and\ \citenamefont {Gross}}]{EMDS16}%
  \BibitemOpen
  \bibfield  {author} {\bibinfo {author} {\bibfnamefont {P.}~\bibnamefont
  {Elliott}}, \bibinfo {author} {\bibfnamefont {T.}~\bibnamefont {Mueller}},
  \bibinfo {author} {\bibfnamefont {J.~K.}\ \bibnamefont {Dewhurst}}, \bibinfo
  {author} {\bibfnamefont {S.}~\bibnamefont {Sharma}}, \ and\ \bibinfo {author}
  {\bibfnamefont {E.~K.~U.}\ \bibnamefont {Gross}},\ }\href@noop {} {\bibfield
  {journal} {\bibinfo  {journal} {Sci Rep}\ }\textbf {\bibinfo {volume} {6}}
  (\bibinfo {year} {2016})}\BibitemShut {NoStop}%
\bibitem [{\citenamefont {Chen}\ \emph {et~al.}(2019)\citenamefont {Chen},
  \citenamefont {Bovensiepen}, \citenamefont {Eschenlohr}, \citenamefont
  {M\"uller}, \citenamefont {Elliott}, \citenamefont {Gross}, \citenamefont
  {Dewhurst},\ and\ \citenamefont {Sharma}}]{CBEM19}%
  \BibitemOpen
  \bibfield  {author} {\bibinfo {author} {\bibfnamefont {J.}~\bibnamefont
  {Chen}}, \bibinfo {author} {\bibfnamefont {U.}~\bibnamefont {Bovensiepen}},
  \bibinfo {author} {\bibfnamefont {A.}~\bibnamefont {Eschenlohr}}, \bibinfo
  {author} {\bibfnamefont {T.}~\bibnamefont {M\"uller}}, \bibinfo {author}
  {\bibfnamefont {P.}~\bibnamefont {Elliott}}, \bibinfo {author} {\bibfnamefont
  {E.~K.~U.}\ \bibnamefont {Gross}}, \bibinfo {author} {\bibfnamefont {J.~K.}\
  \bibnamefont {Dewhurst}}, \ and\ \bibinfo {author} {\bibfnamefont
  {S.}~\bibnamefont {Sharma}},\ }\href {\doibase
  10.1103/PhysRevLett.122.067202} {\bibfield  {journal} {\bibinfo  {journal}
  {Phys. Rev. Lett.}\ }\textbf {\bibinfo {volume} {122}},\ \bibinfo {pages}
  {067202} (\bibinfo {year} {2019})}\BibitemShut {NoStop}%
\bibitem [{\citenamefont {Siegrist}\ \emph {et~al.}(2019)\citenamefont
  {Siegrist}, \citenamefont {Gessner}, \citenamefont {Ossiander}, \citenamefont
  {Denker}, \citenamefont {Chang}, \citenamefont {Schr{\"{o}}der},
  \citenamefont {Guggenmos}, \citenamefont {Cui}, \citenamefont {Walowski},
  \citenamefont {Martens}, \citenamefont {Dewhurst}, \citenamefont
  {Kleineberg}, \citenamefont {M{\"{u}}nzenberg}, \citenamefont {Sharma},\ and\
  \citenamefont {Schultze}}]{Siegrist2019}%
  \BibitemOpen
  \bibfield  {author} {\bibinfo {author} {\bibfnamefont {F.}~\bibnamefont
  {Siegrist}}, \bibinfo {author} {\bibfnamefont {J.~A.}\ \bibnamefont
  {Gessner}}, \bibinfo {author} {\bibfnamefont {M.}~\bibnamefont {Ossiander}},
  \bibinfo {author} {\bibfnamefont {C.}~\bibnamefont {Denker}}, \bibinfo
  {author} {\bibfnamefont {Y.-P.}\ \bibnamefont {Chang}}, \bibinfo {author}
  {\bibfnamefont {M.~C.}\ \bibnamefont {Schr{\"{o}}der}}, \bibinfo {author}
  {\bibfnamefont {A.}~\bibnamefont {Guggenmos}}, \bibinfo {author}
  {\bibfnamefont {Y.}~\bibnamefont {Cui}}, \bibinfo {author} {\bibfnamefont
  {J.}~\bibnamefont {Walowski}}, \bibinfo {author} {\bibfnamefont
  {U.}~\bibnamefont {Martens}}, \bibinfo {author} {\bibfnamefont {J.~K.}\
  \bibnamefont {Dewhurst}}, \bibinfo {author} {\bibfnamefont {U.}~\bibnamefont
  {Kleineberg}}, \bibinfo {author} {\bibfnamefont {M.}~\bibnamefont
  {M{\"{u}}nzenberg}}, \bibinfo {author} {\bibfnamefont {S.}~\bibnamefont
  {Sharma}}, \ and\ \bibinfo {author} {\bibfnamefont {M.}~\bibnamefont
  {Schultze}},\ }\href {\doibase 10.1038/s41586-019-1333-x} {\bibfield
  {journal} {\bibinfo  {journal} {Nature}\ }\textbf {\bibinfo {volume} {571}},\
  \bibinfo {pages} {240} (\bibinfo {year} {2019})}\BibitemShut {NoStop}%
\bibitem [{\citenamefont {Dewhurst}\ \emph
  {et~al.}(2018{\natexlab{a}})\citenamefont {Dewhurst}, \citenamefont
  {Shallcross}, \citenamefont {Gross},\ and\ \citenamefont
  {Sharma}}]{Dewhurst2018}%
  \BibitemOpen
  \bibfield  {author} {\bibinfo {author} {\bibfnamefont {J.~K.}\ \bibnamefont
  {Dewhurst}}, \bibinfo {author} {\bibfnamefont {S.}~\bibnamefont
  {Shallcross}}, \bibinfo {author} {\bibfnamefont {E.~K.~U.}\ \bibnamefont
  {Gross}}, \ and\ \bibinfo {author} {\bibfnamefont {S.}~\bibnamefont
  {Sharma}},\ }\href {\doibase 10.1103/PhysRevApplied.10.044065} {\bibfield
  {journal} {\bibinfo  {journal} {Physical Review Applied}\ }\textbf {\bibinfo
  {volume} {10}},\ \bibinfo {pages} {044065} (\bibinfo {year}
  {2018}{\natexlab{a}})}\BibitemShut {NoStop}%
\bibitem [{\citenamefont {Dewhurst}\ \emph
  {et~al.}(2018{\natexlab{b}})\citenamefont {Dewhurst}, \citenamefont {Sanna},\
  and\ \citenamefont {Sharma}}]{DSS18}%
  \BibitemOpen
  \bibfield  {author} {\bibinfo {author} {\bibfnamefont {J.~K.}\ \bibnamefont
  {Dewhurst}}, \bibinfo {author} {\bibfnamefont {A.}~\bibnamefont {Sanna}}, \
  and\ \bibinfo {author} {\bibfnamefont {S.}~\bibnamefont {Sharma}},\ }\href
  {\doibase 10.1140/epjb/e2018-90146-1} {\bibfield  {journal} {\bibinfo
  {journal} {The European Physical Journal B}\ }\textbf {\bibinfo {volume}
  {91}},\ \bibinfo {pages} {218} (\bibinfo {year}
  {2018}{\natexlab{b}})}\BibitemShut {NoStop}%
\bibitem [{\citenamefont {Willems}\ \emph {et~al.}(2019)\citenamefont
  {Willems}, \citenamefont {Sharma}, \citenamefont {{v. Korff Schmising}},
  \citenamefont {Dewhurst}, \citenamefont {Salemi}, \citenamefont {Schick},
  \citenamefont {Hessing}, \citenamefont {Str{\"{u}}ber}, \citenamefont
  {Engel},\ and\ \citenamefont {Eisebitt}}]{Willems2019}%
  \BibitemOpen
  \bibfield  {author} {\bibinfo {author} {\bibfnamefont {F.}~\bibnamefont
  {Willems}}, \bibinfo {author} {\bibfnamefont {S.}~\bibnamefont {Sharma}},
  \bibinfo {author} {\bibfnamefont {C.}~\bibnamefont {{v. Korff Schmising}}},
  \bibinfo {author} {\bibfnamefont {J.~K.}\ \bibnamefont {Dewhurst}}, \bibinfo
  {author} {\bibfnamefont {L.}~\bibnamefont {Salemi}}, \bibinfo {author}
  {\bibfnamefont {D.}~\bibnamefont {Schick}}, \bibinfo {author} {\bibfnamefont
  {P.}~\bibnamefont {Hessing}}, \bibinfo {author} {\bibfnamefont
  {C.}~\bibnamefont {Str{\"{u}}ber}}, \bibinfo {author} {\bibfnamefont {W.~D.}\
  \bibnamefont {Engel}}, \ and\ \bibinfo {author} {\bibfnamefont
  {S.}~\bibnamefont {Eisebitt}},\ }\href {\doibase
  10.1103/PhysRevLett.122.217202} {\bibfield  {journal} {\bibinfo  {journal}
  {Physical Review Letters}\ }\textbf {\bibinfo {volume} {122}},\ \bibinfo
  {pages} {217202} (\bibinfo {year} {2019})}\BibitemShut {NoStop}%
\bibitem [{\citenamefont {Kampfrath}\ \emph {et~al.}(2013)\citenamefont
  {Kampfrath}, \citenamefont {Battiato}, \citenamefont {Maldonado},
  \citenamefont {Eilers}, \citenamefont {N{\"o}tzold}, \citenamefont
  {M{\"a}hrlein}, \citenamefont {Zbarsky}, \citenamefont {Freimuth},
  \citenamefont {Mokrousov}, \citenamefont {Bl{\"u}gel}, \citenamefont {Wolf},
  \citenamefont {Radu}, \citenamefont {Oppeneer},\ and\ \citenamefont
  {M{\"u}nzenberg}}]{Kampfrath2013}%
  \BibitemOpen
  \bibfield  {author} {\bibinfo {author} {\bibfnamefont {T.}~\bibnamefont
  {Kampfrath}}, \bibinfo {author} {\bibfnamefont {M.}~\bibnamefont {Battiato}},
  \bibinfo {author} {\bibfnamefont {P.}~\bibnamefont {Maldonado}}, \bibinfo
  {author} {\bibfnamefont {G.}~\bibnamefont {Eilers}}, \bibinfo {author}
  {\bibfnamefont {J.}~\bibnamefont {N{\"o}tzold}}, \bibinfo {author}
  {\bibfnamefont {S.}~\bibnamefont {M{\"a}hrlein}}, \bibinfo {author}
  {\bibfnamefont {V.}~\bibnamefont {Zbarsky}}, \bibinfo {author} {\bibfnamefont
  {F.}~\bibnamefont {Freimuth}}, \bibinfo {author} {\bibfnamefont
  {Y.}~\bibnamefont {Mokrousov}}, \bibinfo {author} {\bibfnamefont
  {S.}~\bibnamefont {Bl{\"u}gel}}, \bibinfo {author} {\bibfnamefont
  {M.}~\bibnamefont {Wolf}}, \bibinfo {author} {\bibfnamefont {I.}~\bibnamefont
  {Radu}}, \bibinfo {author} {\bibfnamefont {P.~M.}\ \bibnamefont {Oppeneer}},
  \ and\ \bibinfo {author} {\bibfnamefont {M.}~\bibnamefont {M{\"u}nzenberg}},\
  }\href {https://doi.org/10.1038/nnano.2013.43} {\bibfield  {journal}
  {\bibinfo  {journal} {Nature Nanotechnology}\ }\textbf {\bibinfo {volume}
  {8}},\ \bibinfo {pages} {256 EP } (\bibinfo {year} {2013})}\BibitemShut
  {NoStop}%
\bibitem [{\citenamefont {Huisman}\ \emph {et~al.}(2016)\citenamefont
  {Huisman}, \citenamefont {Mikhaylovskiy}, \citenamefont {Costa},
  \citenamefont {Freimuth}, \citenamefont {Paz}, \citenamefont {Ventura},
  \citenamefont {Freitas}, \citenamefont {Bl{\"u}gel}, \citenamefont
  {Mokrousov}, \citenamefont {Rasing},\ and\ \citenamefont
  {Kimel}}]{Huisman2016}%
  \BibitemOpen
  \bibfield  {author} {\bibinfo {author} {\bibfnamefont {T.~J.}\ \bibnamefont
  {Huisman}}, \bibinfo {author} {\bibfnamefont {R.~V.}\ \bibnamefont
  {Mikhaylovskiy}}, \bibinfo {author} {\bibfnamefont {J.~D.}\ \bibnamefont
  {Costa}}, \bibinfo {author} {\bibfnamefont {F.}~\bibnamefont {Freimuth}},
  \bibinfo {author} {\bibfnamefont {E.}~\bibnamefont {Paz}}, \bibinfo {author}
  {\bibfnamefont {J.}~\bibnamefont {Ventura}}, \bibinfo {author} {\bibfnamefont
  {P.~P.}\ \bibnamefont {Freitas}}, \bibinfo {author} {\bibfnamefont
  {S.}~\bibnamefont {Bl{\"u}gel}}, \bibinfo {author} {\bibfnamefont
  {Y.}~\bibnamefont {Mokrousov}}, \bibinfo {author} {\bibfnamefont
  {T.}~\bibnamefont {Rasing}}, \ and\ \bibinfo {author} {\bibfnamefont {A.~V.}\
  \bibnamefont {Kimel}},\ }\href {https://doi.org/10.1038/nnano.2015.331}
  {\bibfield  {journal} {\bibinfo  {journal} {Nature Nanotechnology}\ }\textbf
  {\bibinfo {volume} {11}},\ \bibinfo {pages} {455 EP } (\bibinfo {year}
  {2016})}\BibitemShut {NoStop}%
\bibitem [{\citenamefont {Lambert}\ \emph {et~al.}(2014)\citenamefont
  {Lambert}, \citenamefont {Mangin}, \citenamefont {Varaprasad}, \citenamefont
  {Takahashi}, \citenamefont {Hehn}, \citenamefont {Cinchetti}, \citenamefont
  {Malinowski}, \citenamefont {Hono}, \citenamefont {Fainman}, \citenamefont
  {Aeschlimann},\ and\ \citenamefont {Fullerton}}]{Lambert2014}%
  \BibitemOpen
  \bibfield  {author} {\bibinfo {author} {\bibfnamefont {C.~H.}\ \bibnamefont
  {Lambert}}, \bibinfo {author} {\bibfnamefont {S.}~\bibnamefont {Mangin}},
  \bibinfo {author} {\bibfnamefont {B.~S.~S.}\ \bibnamefont {Varaprasad}},
  \bibinfo {author} {\bibfnamefont {Y.~K.}\ \bibnamefont {Takahashi}}, \bibinfo
  {author} {\bibfnamefont {M.}~\bibnamefont {Hehn}}, \bibinfo {author}
  {\bibfnamefont {M.}~\bibnamefont {Cinchetti}}, \bibinfo {author}
  {\bibfnamefont {G.}~\bibnamefont {Malinowski}}, \bibinfo {author}
  {\bibfnamefont {K.}~\bibnamefont {Hono}}, \bibinfo {author} {\bibfnamefont
  {Y.}~\bibnamefont {Fainman}}, \bibinfo {author} {\bibfnamefont
  {M.}~\bibnamefont {Aeschlimann}}, \ and\ \bibinfo {author} {\bibfnamefont
  {E.~E.}\ \bibnamefont {Fullerton}},\ }\href {\doibase
  10.1126/science.1253493} {\bibfield  {journal} {\bibinfo  {journal}
  {Science}\ }\textbf {\bibinfo {volume} {345}},\ \bibinfo {pages} {1337}
  (\bibinfo {year} {2014})},\ \Eprint {http://arxiv.org/abs/1403.0784}
  {arXiv:1403.0784} \BibitemShut {NoStop}%
\bibitem [{\citenamefont {Willems}\ \emph
  {et~al.}(2015{\natexlab{a}})\citenamefont {Willems}, \citenamefont {Smeenk},
  \citenamefont {Zhavoronkov}, \citenamefont {Kornilov}, \citenamefont {Radu},
  \citenamefont {Schmidbauer}, \citenamefont {Hanke}, \citenamefont {{Von Korff
  Schmising}}, \citenamefont {Vrakking},\ and\ \citenamefont
  {Eisebitt}}]{Willems2015}%
  \BibitemOpen
  \bibfield  {author} {\bibinfo {author} {\bibfnamefont {F.}~\bibnamefont
  {Willems}}, \bibinfo {author} {\bibfnamefont {C.~T.}\ \bibnamefont {Smeenk}},
  \bibinfo {author} {\bibfnamefont {N.}~\bibnamefont {Zhavoronkov}}, \bibinfo
  {author} {\bibfnamefont {O.}~\bibnamefont {Kornilov}}, \bibinfo {author}
  {\bibfnamefont {I.}~\bibnamefont {Radu}}, \bibinfo {author} {\bibfnamefont
  {M.}~\bibnamefont {Schmidbauer}}, \bibinfo {author} {\bibfnamefont
  {M.}~\bibnamefont {Hanke}}, \bibinfo {author} {\bibfnamefont
  {C.}~\bibnamefont {{Von Korff Schmising}}}, \bibinfo {author} {\bibfnamefont
  {M.~J.}\ \bibnamefont {Vrakking}}, \ and\ \bibinfo {author} {\bibfnamefont
  {S.}~\bibnamefont {Eisebitt}},\ }\href {\doibase 10.1103/PhysRevB.92.220405}
  {\bibfield  {journal} {\bibinfo  {journal} {Physical Review B - Condensed
  Matter and Materials Physics}\ }\textbf {\bibinfo {volume} {92}},\ \bibinfo
  {pages} {1} (\bibinfo {year} {2015}{\natexlab{a}})}\BibitemShut {NoStop}%
\bibitem [{\citenamefont {Dewhurst}\ \emph
  {et~al.}(2018{\natexlab{c}})\citenamefont {Dewhurst}, \citenamefont
  {Elliott}, \citenamefont {Shallcross}, \citenamefont {Gross},\ and\
  \citenamefont {Sharma}}]{DESG18}%
  \BibitemOpen
  \bibfield  {author} {\bibinfo {author} {\bibfnamefont {J.~K.}\ \bibnamefont
  {Dewhurst}}, \bibinfo {author} {\bibfnamefont {P.}~\bibnamefont {Elliott}},
  \bibinfo {author} {\bibfnamefont {S.}~\bibnamefont {Shallcross}}, \bibinfo
  {author} {\bibfnamefont {E.~K.~U.}\ \bibnamefont {Gross}}, \ and\ \bibinfo
  {author} {\bibfnamefont {S.}~\bibnamefont {Sharma}},\ }\href {\doibase
  10.1021/acs.nanolett.7b05118} {\bibfield  {journal} {\bibinfo  {journal}
  {Nano Letters}\ }\textbf {\bibinfo {volume} {18}},\ \bibinfo {pages} {1842}
  (\bibinfo {year} {2018}{\natexlab{c}})},\ \bibinfo {note} {pMID: 29424230},\
  \Eprint {http://arxiv.org/abs/https://doi.org/10.1021/acs.nanolett.7b05118}
  {https://doi.org/10.1021/acs.nanolett.7b05118} \BibitemShut {NoStop}%
\bibitem [{\citenamefont {Shishidou}\ \emph {et~al.}(1997)\citenamefont
  {Shishidou}, \citenamefont {Imada}, \citenamefont {Muro}, \citenamefont
  {Oda}, \citenamefont {Kimura}, \citenamefont {Suga}, \citenamefont
  {Miyahara}, \citenamefont {Kanomata},\ and\ \citenamefont
  {Kaneko}}]{Shishidou1997a}%
  \BibitemOpen
  \bibfield  {author} {\bibinfo {author} {\bibfnamefont {T.}~\bibnamefont
  {Shishidou}}, \bibinfo {author} {\bibfnamefont {S.}~\bibnamefont {Imada}},
  \bibinfo {author} {\bibfnamefont {T.}~\bibnamefont {Muro}}, \bibinfo {author}
  {\bibfnamefont {F.}~\bibnamefont {Oda}}, \bibinfo {author} {\bibfnamefont
  {A.}~\bibnamefont {Kimura}}, \bibinfo {author} {\bibfnamefont
  {S.}~\bibnamefont {Suga}}, \bibinfo {author} {\bibfnamefont {T.}~\bibnamefont
  {Miyahara}}, \bibinfo {author} {\bibfnamefont {T.}~\bibnamefont {Kanomata}},
  \ and\ \bibinfo {author} {\bibfnamefont {T.}~\bibnamefont {Kaneko}},\ }\href
  {\doibase 10.1103/PhysRevB.55.3749} {\bibfield  {journal} {\bibinfo
  {journal} {Phys. Rev. B}\ }\textbf {\bibinfo {volume} {55}},\ \bibinfo
  {pages} {3749} (\bibinfo {year} {1997})}\BibitemShut {NoStop}%
\bibitem [{\citenamefont {Nakajima}\ \emph {et~al.}(1998)\citenamefont
  {Nakajima}, \citenamefont {Koide}, \citenamefont {Shidara}, \citenamefont
  {Miyauchi}, \citenamefont {Fukutani}, \citenamefont {Fujimori}, \citenamefont
  {Iio}, \citenamefont {Katayama}, \citenamefont {N{\'{y}}vlt},\ and\
  \citenamefont {Suzuki}}]{Nakajima2002}%
  \BibitemOpen
  \bibfield  {author} {\bibinfo {author} {\bibfnamefont {N.}~\bibnamefont
  {Nakajima}}, \bibinfo {author} {\bibfnamefont {T.}~\bibnamefont {Koide}},
  \bibinfo {author} {\bibfnamefont {T.}~\bibnamefont {Shidara}}, \bibinfo
  {author} {\bibfnamefont {H.}~\bibnamefont {Miyauchi}}, \bibinfo {author}
  {\bibfnamefont {H.}~\bibnamefont {Fukutani}}, \bibinfo {author}
  {\bibfnamefont {A.}~\bibnamefont {Fujimori}}, \bibinfo {author}
  {\bibfnamefont {K.}~\bibnamefont {Iio}}, \bibinfo {author} {\bibfnamefont
  {T.}~\bibnamefont {Katayama}}, \bibinfo {author} {\bibfnamefont
  {M.}~\bibnamefont {N{\'{y}}vlt}}, \ and\ \bibinfo {author} {\bibfnamefont
  {Y.}~\bibnamefont {Suzuki}},\ }\href {\doibase 10.1103/PhysRevLett.81.5229}
  {\bibfield  {journal} {\bibinfo  {journal} {Physical Review Letters}\
  }\textbf {\bibinfo {volume} {81}},\ \bibinfo {pages} {5229} (\bibinfo {year}
  {1998})}\BibitemShut {NoStop}%
\bibitem [{\citenamefont {Willems}\ \emph {et~al.}(2017)\citenamefont
  {Willems}, \citenamefont {von Korff~Schmising}, \citenamefont {Weder},
  \citenamefont {G\"unther}, \citenamefont {Schneider}, \citenamefont {Pfau},
  \citenamefont {Meise}, \citenamefont {Guehrs}, \citenamefont {Geilhufe},
  \citenamefont {Merhe}, \citenamefont {Jal}, \citenamefont {Vodungbo},
  \citenamefont {Lning}, \citenamefont {Mahieu}, \citenamefont {Capotondi},
  \citenamefont {Pedersoli}, \citenamefont {Gauthier}, \citenamefont
  {Manfredda},\ and\ \citenamefont {Eisebitt}}]{Willems2017}%
  \BibitemOpen
  \bibfield  {author} {\bibinfo {author} {\bibfnamefont {F.}~\bibnamefont
  {Willems}}, \bibinfo {author} {\bibfnamefont {C.}~\bibnamefont {von
  Korff~Schmising}}, \bibinfo {author} {\bibfnamefont {D.}~\bibnamefont
  {Weder}}, \bibinfo {author} {\bibfnamefont {C.~M.}\ \bibnamefont
  {G\"unther}}, \bibinfo {author} {\bibfnamefont {M.}~\bibnamefont
  {Schneider}}, \bibinfo {author} {\bibfnamefont {B.}~\bibnamefont {Pfau}},
  \bibinfo {author} {\bibfnamefont {S.}~\bibnamefont {Meise}}, \bibinfo
  {author} {\bibfnamefont {E.}~\bibnamefont {Guehrs}}, \bibinfo {author}
  {\bibfnamefont {J.}~\bibnamefont {Geilhufe}}, \bibinfo {author}
  {\bibfnamefont {A.~E.~D.}\ \bibnamefont {Merhe}}, \bibinfo {author}
  {\bibfnamefont {E.}~\bibnamefont {Jal}}, \bibinfo {author} {\bibfnamefont
  {B.}~\bibnamefont {Vodungbo}}, \bibinfo {author} {\bibfnamefont
  {J.}~\bibnamefont {Lning}}, \bibinfo {author} {\bibfnamefont
  {B.}~\bibnamefont {Mahieu}}, \bibinfo {author} {\bibfnamefont
  {F.}~\bibnamefont {Capotondi}}, \bibinfo {author} {\bibfnamefont
  {E.}~\bibnamefont {Pedersoli}}, \bibinfo {author} {\bibfnamefont
  {D.}~\bibnamefont {Gauthier}}, \bibinfo {author} {\bibfnamefont
  {M.}~\bibnamefont {Manfredda}}, \ and\ \bibinfo {author} {\bibfnamefont
  {S.}~\bibnamefont {Eisebitt}},\ }\href {\doibase 10.1063/1.4976004}
  {\bibfield  {journal} {\bibinfo  {journal} {Structural Dynamics}\ }\textbf
  {\bibinfo {volume} {4}},\ \bibinfo {pages} {014301} (\bibinfo {year}
  {2017})}\BibitemShut {NoStop}%
\bibitem [{\citenamefont {Oppeneer}\ \emph {et~al.}(1992)\citenamefont
  {Oppeneer}, \citenamefont {Maurer}, \citenamefont {Sticht},\ and\
  \citenamefont {K\"ubler}}]{OMSK92}%
  \BibitemOpen
  \bibfield  {author} {\bibinfo {author} {\bibfnamefont {P.~M.}\ \bibnamefont
  {Oppeneer}}, \bibinfo {author} {\bibfnamefont {T.}~\bibnamefont {Maurer}},
  \bibinfo {author} {\bibfnamefont {J.}~\bibnamefont {Sticht}}, \ and\ \bibinfo
  {author} {\bibfnamefont {J.}~\bibnamefont {K\"ubler}},\ }\href {\doibase
  10.1103/PhysRevB.45.10924} {\bibfield  {journal} {\bibinfo  {journal} {Phys.
  Rev. B}\ }\textbf {\bibinfo {volume} {45}},\ \bibinfo {pages} {10924}
  (\bibinfo {year} {1992})}\BibitemShut {NoStop}%
\bibitem [{\citenamefont {{von Korff Schmising}}\ \emph
  {et~al.}(2017)\citenamefont {{von Korff Schmising}}, \citenamefont {Weder},
  \citenamefont {Noll}, \citenamefont {Pfau}, \citenamefont {Hennecke},
  \citenamefont {Str{\"{u}}ber}, \citenamefont {Radu}, \citenamefont
  {Schneider}, \citenamefont {Staeck}, \citenamefont {G{\"{u}}nther},
  \citenamefont {L{\"{u}}ning}, \citenamefont {Merhe}, \citenamefont {Buck},
  \citenamefont {Hartmann}, \citenamefont {Viefhaus}, \citenamefont {Treusch},\
  and\ \citenamefont {Eisebitt}}]{vonKorffSchmising2017}%
  \BibitemOpen
  \bibfield  {author} {\bibinfo {author} {\bibfnamefont {C.}~\bibnamefont {{von
  Korff Schmising}}}, \bibinfo {author} {\bibfnamefont {D.}~\bibnamefont
  {Weder}}, \bibinfo {author} {\bibfnamefont {T.}~\bibnamefont {Noll}},
  \bibinfo {author} {\bibfnamefont {B.}~\bibnamefont {Pfau}}, \bibinfo {author}
  {\bibfnamefont {M.}~\bibnamefont {Hennecke}}, \bibinfo {author}
  {\bibfnamefont {C.}~\bibnamefont {Str{\"{u}}ber}}, \bibinfo {author}
  {\bibfnamefont {I.}~\bibnamefont {Radu}}, \bibinfo {author} {\bibfnamefont
  {M.}~\bibnamefont {Schneider}}, \bibinfo {author} {\bibfnamefont
  {S.}~\bibnamefont {Staeck}}, \bibinfo {author} {\bibfnamefont {C.~M.}\
  \bibnamefont {G{\"{u}}nther}}, \bibinfo {author} {\bibfnamefont
  {J.}~\bibnamefont {L{\"{u}}ning}}, \bibinfo {author} {\bibfnamefont
  {A.~E.~D.}\ \bibnamefont {Merhe}}, \bibinfo {author} {\bibfnamefont
  {J.}~\bibnamefont {Buck}}, \bibinfo {author} {\bibfnamefont {G.}~\bibnamefont
  {Hartmann}}, \bibinfo {author} {\bibfnamefont {J.}~\bibnamefont {Viefhaus}},
  \bibinfo {author} {\bibfnamefont {R.}~\bibnamefont {Treusch}}, \ and\
  \bibinfo {author} {\bibfnamefont {S.}~\bibnamefont {Eisebitt}},\ }\href
  {\doibase 10.1063/1.4983056} {\bibfield  {journal} {\bibinfo  {journal}
  {Review of Scientific Instruments}\ }\textbf {\bibinfo {volume} {88}},\
  \bibinfo {pages} {053903} (\bibinfo {year} {2017})}\BibitemShut {NoStop}%
\bibitem [{\citenamefont {Willems}\ \emph
  {et~al.}(2015{\natexlab{b}})\citenamefont {Willems}, \citenamefont {Smeenk},
  \citenamefont {Zhavoronkov}, \citenamefont {Kornilov}, \citenamefont {Radu},
  \citenamefont {Schmidbauer}, \citenamefont {Hanke}, \citenamefont {von
  Korff~Schmising}, \citenamefont {Vrakking},\ and\ \citenamefont
  {Eisebitt}}]{Willems15}%
  \BibitemOpen
  \bibfield  {author} {\bibinfo {author} {\bibfnamefont {F.}~\bibnamefont
  {Willems}}, \bibinfo {author} {\bibfnamefont {C.~T.~L.}\ \bibnamefont
  {Smeenk}}, \bibinfo {author} {\bibfnamefont {N.}~\bibnamefont {Zhavoronkov}},
  \bibinfo {author} {\bibfnamefont {O.}~\bibnamefont {Kornilov}}, \bibinfo
  {author} {\bibfnamefont {I.}~\bibnamefont {Radu}}, \bibinfo {author}
  {\bibfnamefont {M.}~\bibnamefont {Schmidbauer}}, \bibinfo {author}
  {\bibfnamefont {M.}~\bibnamefont {Hanke}}, \bibinfo {author} {\bibfnamefont
  {C.}~\bibnamefont {von Korff~Schmising}}, \bibinfo {author} {\bibfnamefont
  {M.~J.~J.}\ \bibnamefont {Vrakking}}, \ and\ \bibinfo {author} {\bibfnamefont
  {S.}~\bibnamefont {Eisebitt}},\ }\href {\doibase 10.1103/PhysRevB.92.220405}
  {\bibfield  {journal} {\bibinfo  {journal} {Phys. Rev. B}\ }\textbf {\bibinfo
  {volume} {92}},\ \bibinfo {pages} {220405} (\bibinfo {year}
  {2015}{\natexlab{b}})}\BibitemShut {NoStop}%
\bibitem [{\citenamefont {Hofherr}\ \emph {et~al.}(2018)\citenamefont
  {Hofherr}, \citenamefont {Moretti}, \citenamefont {Shim}, \citenamefont
  {H\"auser}, \citenamefont {Safonova}, \citenamefont {Stiehl}, \citenamefont
  {Ali}, \citenamefont {Sakshath}, \citenamefont {Kim}, \citenamefont {Kim},
  \citenamefont {Kim}, \citenamefont {Hong}, \citenamefont {Kapteyn},
  \citenamefont {Murnane}, \citenamefont {Cinchetti}, \citenamefont {Steil},
  \citenamefont {Mathias}, \citenamefont {Stadtm\"uller}, \citenamefont
  {Albrecht}, \citenamefont {Kim}, \citenamefont {Nowak},\ and\ \citenamefont
  {Aeschlimann}}]{Hofherr18}%
  \BibitemOpen
  \bibfield  {author} {\bibinfo {author} {\bibfnamefont {M.}~\bibnamefont
  {Hofherr}}, \bibinfo {author} {\bibfnamefont {S.}~\bibnamefont {Moretti}},
  \bibinfo {author} {\bibfnamefont {J.}~\bibnamefont {Shim}}, \bibinfo {author}
  {\bibfnamefont {S.}~\bibnamefont {H\"auser}}, \bibinfo {author}
  {\bibfnamefont {N.~Y.}\ \bibnamefont {Safonova}}, \bibinfo {author}
  {\bibfnamefont {M.}~\bibnamefont {Stiehl}}, \bibinfo {author} {\bibfnamefont
  {A.}~\bibnamefont {Ali}}, \bibinfo {author} {\bibfnamefont {S.}~\bibnamefont
  {Sakshath}}, \bibinfo {author} {\bibfnamefont {J.~W.}\ \bibnamefont {Kim}},
  \bibinfo {author} {\bibfnamefont {D.~H.}\ \bibnamefont {Kim}}, \bibinfo
  {author} {\bibfnamefont {H.~J.}\ \bibnamefont {Kim}}, \bibinfo {author}
  {\bibfnamefont {J.~I.}\ \bibnamefont {Hong}}, \bibinfo {author}
  {\bibfnamefont {H.~C.}\ \bibnamefont {Kapteyn}}, \bibinfo {author}
  {\bibfnamefont {M.~M.}\ \bibnamefont {Murnane}}, \bibinfo {author}
  {\bibfnamefont {M.}~\bibnamefont {Cinchetti}}, \bibinfo {author}
  {\bibfnamefont {D.}~\bibnamefont {Steil}}, \bibinfo {author} {\bibfnamefont
  {S.}~\bibnamefont {Mathias}}, \bibinfo {author} {\bibfnamefont
  {B.}~\bibnamefont {Stadtm\"uller}}, \bibinfo {author} {\bibfnamefont
  {M.}~\bibnamefont {Albrecht}}, \bibinfo {author} {\bibfnamefont {D.~E.}\
  \bibnamefont {Kim}}, \bibinfo {author} {\bibfnamefont {U.}~\bibnamefont
  {Nowak}}, \ and\ \bibinfo {author} {\bibfnamefont {M.}~\bibnamefont
  {Aeschlimann}},\ }\href {\doibase 10.1103/PhysRevB.98.174419} {\bibfield
  {journal} {\bibinfo  {journal} {Phys. Rev. B}\ }\textbf {\bibinfo {volume}
  {98}},\ \bibinfo {pages} {174419} (\bibinfo {year} {2018})}\BibitemShut
  {NoStop}%
\bibitem [{\citenamefont {La-O-Vorakiat}\ \emph {et~al.}(2009)\citenamefont
  {La-O-Vorakiat}, \citenamefont {Siemens}, \citenamefont {Murnane},
  \citenamefont {Kapteyn}, \citenamefont {Mathias}, \citenamefont
  {Aeschlimann}, \citenamefont {Grychtol}, \citenamefont {Adam}, \citenamefont
  {Schneider}, \citenamefont {Shaw}, \citenamefont {Nembach},\ and\
  \citenamefont {Silva}}]{Vorakiat2009}%
  \BibitemOpen
  \bibfield  {author} {\bibinfo {author} {\bibfnamefont {C.}~\bibnamefont
  {La-O-Vorakiat}}, \bibinfo {author} {\bibfnamefont {M.}~\bibnamefont
  {Siemens}}, \bibinfo {author} {\bibfnamefont {M.~M.}\ \bibnamefont
  {Murnane}}, \bibinfo {author} {\bibfnamefont {H.~C.}\ \bibnamefont
  {Kapteyn}}, \bibinfo {author} {\bibfnamefont {S.}~\bibnamefont {Mathias}},
  \bibinfo {author} {\bibfnamefont {M.}~\bibnamefont {Aeschlimann}}, \bibinfo
  {author} {\bibfnamefont {P.}~\bibnamefont {Grychtol}}, \bibinfo {author}
  {\bibfnamefont {R.}~\bibnamefont {Adam}}, \bibinfo {author} {\bibfnamefont
  {C.~M.}\ \bibnamefont {Schneider}}, \bibinfo {author} {\bibfnamefont {J.~M.}\
  \bibnamefont {Shaw}}, \bibinfo {author} {\bibfnamefont {H.}~\bibnamefont
  {Nembach}}, \ and\ \bibinfo {author} {\bibfnamefont {T.~J.}\ \bibnamefont
  {Silva}},\ }\href {\doibase 10.1103/PhysRevLett.103.257402} {\bibfield
  {journal} {\bibinfo  {journal} {Physical Review Letters}\ }\textbf {\bibinfo
  {volume} {103}},\ \bibinfo {pages} {257402} (\bibinfo {year}
  {2009})}\BibitemShut {NoStop}%
\bibitem [{\citenamefont {Mathias}\ \emph {et~al.}(2012)\citenamefont
  {Mathias}, \citenamefont {La-O-Vorakiat}, \citenamefont {Grychtol},
  \citenamefont {Granitzka}, \citenamefont {Turgut}, \citenamefont {Shaw},
  \citenamefont {Adam}, \citenamefont {Nembach}, \citenamefont {Siemens},
  \citenamefont {Eich}, \citenamefont {Schneider}, \citenamefont {Silva},
  \citenamefont {Aeschlimann}, \citenamefont {Murnane},\ and\ \citenamefont
  {Kapteyn}}]{Mathias2012}%
  \BibitemOpen
  \bibfield  {author} {\bibinfo {author} {\bibfnamefont {S.}~\bibnamefont
  {Mathias}}, \bibinfo {author} {\bibfnamefont {C.}~\bibnamefont
  {La-O-Vorakiat}}, \bibinfo {author} {\bibfnamefont {P.}~\bibnamefont
  {Grychtol}}, \bibinfo {author} {\bibfnamefont {P.}~\bibnamefont {Granitzka}},
  \bibinfo {author} {\bibfnamefont {E.}~\bibnamefont {Turgut}}, \bibinfo
  {author} {\bibfnamefont {J.~M.}\ \bibnamefont {Shaw}}, \bibinfo {author}
  {\bibfnamefont {R.}~\bibnamefont {Adam}}, \bibinfo {author} {\bibfnamefont
  {H.~T.}\ \bibnamefont {Nembach}}, \bibinfo {author} {\bibfnamefont {M.~E.}\
  \bibnamefont {Siemens}}, \bibinfo {author} {\bibfnamefont {S.}~\bibnamefont
  {Eich}}, \bibinfo {author} {\bibfnamefont {C.~M.}\ \bibnamefont {Schneider}},
  \bibinfo {author} {\bibfnamefont {T.~J.}\ \bibnamefont {Silva}}, \bibinfo
  {author} {\bibfnamefont {M.}~\bibnamefont {Aeschlimann}}, \bibinfo {author}
  {\bibfnamefont {M.~M.}\ \bibnamefont {Murnane}}, \ and\ \bibinfo {author}
  {\bibfnamefont {H.~C.}\ \bibnamefont {Kapteyn}},\ }\href {\doibase
  10.1073/pnas.1201371109} {\bibfield  {journal} {\bibinfo  {journal}
  {Proceedings of the National Academy of Sciences}\ }\textbf {\bibinfo
  {volume} {109}},\ \bibinfo {pages} {4792} (\bibinfo {year}
  {2012})}\BibitemShut {NoStop}%
\bibitem [{\citenamefont {Gang}\ \emph {et~al.}(2018)\citenamefont {Gang},
  \citenamefont {Adam}, \citenamefont {Pl{\"{o}}tzing}, \citenamefont {von
  Witzleben}, \citenamefont {Weier}, \citenamefont {Parlak}, \citenamefont
  {B{\"{u}}rgler}, \citenamefont {Schneider}, \citenamefont {Rusz},
  \citenamefont {Maldonado},\ and\ \citenamefont {Oppeneer}}]{Gang2018}%
  \BibitemOpen
  \bibfield  {author} {\bibinfo {author} {\bibfnamefont {S.-g.}\ \bibnamefont
  {Gang}}, \bibinfo {author} {\bibfnamefont {R.}~\bibnamefont {Adam}}, \bibinfo
  {author} {\bibfnamefont {M.}~\bibnamefont {Pl{\"{o}}tzing}}, \bibinfo
  {author} {\bibfnamefont {M.}~\bibnamefont {von Witzleben}}, \bibinfo {author}
  {\bibfnamefont {C.}~\bibnamefont {Weier}}, \bibinfo {author} {\bibfnamefont
  {U.}~\bibnamefont {Parlak}}, \bibinfo {author} {\bibfnamefont {D.~E.}\
  \bibnamefont {B{\"{u}}rgler}}, \bibinfo {author} {\bibfnamefont {C.~M.}\
  \bibnamefont {Schneider}}, \bibinfo {author} {\bibfnamefont {J.}~\bibnamefont
  {Rusz}}, \bibinfo {author} {\bibfnamefont {P.}~\bibnamefont {Maldonado}}, \
  and\ \bibinfo {author} {\bibfnamefont {P.~M.}\ \bibnamefont {Oppeneer}},\
  }\href {\doibase 10.1103/PhysRevB.97.064412} {\bibfield  {journal} {\bibinfo
  {journal} {Physical Review B}\ }\textbf {\bibinfo {volume} {97}},\ \bibinfo
  {pages} {064412} (\bibinfo {year} {2018})}\BibitemShut {NoStop}%
\bibitem [{\citenamefont {G{\"{u}}nther}\ \emph {et~al.}(2014)\citenamefont
  {G{\"{u}}nther}, \citenamefont {Spezzani}, \citenamefont {Ciprian},
  \citenamefont {Grazioli}, \citenamefont {Ressel}, \citenamefont {Coreno},
  \citenamefont {Poletto}, \citenamefont {Miotti}, \citenamefont {Sacchi},
  \citenamefont {Panaccione}, \citenamefont {Uhl{\'{i}}r}, \citenamefont
  {Fullerton}, \citenamefont {{De Ninno}},\ and\ \citenamefont
  {Back}}]{Gunther2014}%
  \BibitemOpen
  \bibfield  {author} {\bibinfo {author} {\bibfnamefont {S.}~\bibnamefont
  {G{\"{u}}nther}}, \bibinfo {author} {\bibfnamefont {C.}~\bibnamefont
  {Spezzani}}, \bibinfo {author} {\bibfnamefont {R.}~\bibnamefont {Ciprian}},
  \bibinfo {author} {\bibfnamefont {C.}~\bibnamefont {Grazioli}}, \bibinfo
  {author} {\bibfnamefont {B.}~\bibnamefont {Ressel}}, \bibinfo {author}
  {\bibfnamefont {M.}~\bibnamefont {Coreno}}, \bibinfo {author} {\bibfnamefont
  {L.}~\bibnamefont {Poletto}}, \bibinfo {author} {\bibfnamefont
  {P.}~\bibnamefont {Miotti}}, \bibinfo {author} {\bibfnamefont
  {M.}~\bibnamefont {Sacchi}}, \bibinfo {author} {\bibfnamefont
  {G.}~\bibnamefont {Panaccione}}, \bibinfo {author} {\bibfnamefont
  {V.}~\bibnamefont {Uhl{\'{i}}r}}, \bibinfo {author} {\bibfnamefont {E.~E.}\
  \bibnamefont {Fullerton}}, \bibinfo {author} {\bibfnamefont {G.}~\bibnamefont
  {{De Ninno}}}, \ and\ \bibinfo {author} {\bibfnamefont {C.~H.}\ \bibnamefont
  {Back}},\ }\href {\doibase 10.1103/PhysRevB.90.180407} {\bibfield  {journal}
  {\bibinfo  {journal} {Physical Review B - Condensed Matter and Materials
  Physics}\ }\textbf {\bibinfo {volume} {90}},\ \bibinfo {pages} {1} (\bibinfo
  {year} {2014})}\BibitemShut {NoStop}%
\bibitem [{\citenamefont {Zhang}\ and\ \citenamefont {Huebner}(2000)}]{ZH00}%
  \BibitemOpen
  \bibfield  {author} {\bibinfo {author} {\bibfnamefont {G.~P.}\ \bibnamefont
  {Zhang}}\ and\ \bibinfo {author} {\bibfnamefont {W.}~\bibnamefont
  {Huebner}},\ }\href {\doibase 10.1103/PhysRevLett.85.3025} {\bibfield
  {journal} {\bibinfo  {journal} {Phys. Rev. Lett.}\ }\textbf {\bibinfo
  {volume} {85}},\ \bibinfo {pages} {3025} (\bibinfo {year}
  {2000})}\BibitemShut {NoStop}%
\bibitem [{\citenamefont {T\"ows}\ and\ \citenamefont {Pastor}(2015)}]{TP15}%
  \BibitemOpen
  \bibfield  {author} {\bibinfo {author} {\bibfnamefont {W.}~\bibnamefont
  {T\"ows}}\ and\ \bibinfo {author} {\bibfnamefont {G.~M.}\ \bibnamefont
  {Pastor}},\ }\href {\doibase 10.1103/PhysRevLett.115.217204} {\bibfield
  {journal} {\bibinfo  {journal} {Phys. Rev. Lett.}\ }\textbf {\bibinfo
  {volume} {115}},\ \bibinfo {pages} {217204} (\bibinfo {year}
  {2015})}\BibitemShut {NoStop}%
\bibitem [{\citenamefont {Kuiper}\ \emph {et~al.}(2014)\citenamefont {Kuiper},
  \citenamefont {Roth}, \citenamefont {Schellekens}, \citenamefont {Schmitt},
  \citenamefont {Koopmans}, \citenamefont {Cinchetti},\ and\ \citenamefont
  {Aeschlimann}}]{Kuiper14}%
  \BibitemOpen
  \bibfield  {author} {\bibinfo {author} {\bibfnamefont {K.~C.}\ \bibnamefont
  {Kuiper}}, \bibinfo {author} {\bibfnamefont {T.}~\bibnamefont {Roth}},
  \bibinfo {author} {\bibfnamefont {A.~J.}\ \bibnamefont {Schellekens}},
  \bibinfo {author} {\bibfnamefont {O.}~\bibnamefont {Schmitt}}, \bibinfo
  {author} {\bibfnamefont {B.}~\bibnamefont {Koopmans}}, \bibinfo {author}
  {\bibfnamefont {M.}~\bibnamefont {Cinchetti}}, \ and\ \bibinfo {author}
  {\bibfnamefont {M.}~\bibnamefont {Aeschlimann}},\ }\href {\doibase
  10.1063/1.4902069} {\bibfield  {journal} {\bibinfo  {journal} {Applied
  Physics Letters}\ }\textbf {\bibinfo {volume} {105}},\ \bibinfo {pages}
  {202402} (\bibinfo {year} {2014})}\BibitemShut {NoStop}%
\bibitem [{\citenamefont {Runge}\ and\ \citenamefont
  {Gross}(1984{\natexlab{b}})}]{Runge1984}%
  \BibitemOpen
  \bibfield  {author} {\bibinfo {author} {\bibfnamefont {E.}~\bibnamefont
  {Runge}}\ and\ \bibinfo {author} {\bibfnamefont {E.~K.~U.}\ \bibnamefont
  {Gross}},\ }\href {\doibase 10.1103/PhysRevLett.52.997} {\bibfield  {journal}
  {\bibinfo  {journal} {Physical Review Letters}\ }\textbf {\bibinfo {volume}
  {52}},\ \bibinfo {pages} {997} (\bibinfo {year}
  {1984}{\natexlab{b}})}\BibitemShut {NoStop}%
\bibitem [{\citenamefont {Hohenberg}\ and\ \citenamefont {Kohn}(1964)}]{HK64}%
  \BibitemOpen
  \bibfield  {author} {\bibinfo {author} {\bibfnamefont {P.}~\bibnamefont
  {Hohenberg}}\ and\ \bibinfo {author} {\bibfnamefont {W.}~\bibnamefont
  {Kohn}},\ }\href@noop {} {\bibfield  {journal} {\bibinfo  {journal} {Phys.
  Rev.}\ }\textbf {\bibinfo {volume} {136}},\ \bibinfo {pages} {B 864}
  (\bibinfo {year} {1964})}\BibitemShut {NoStop}%
\bibitem [{\citenamefont {Singh}(1994)}]{singh}%
  \BibitemOpen
  \bibfield  {author} {\bibinfo {author} {\bibfnamefont {D.~J.}\ \bibnamefont
  {Singh}},\ }\href@noop {} {\emph {\bibinfo {title} {Planewaves
  Pseudopotentials and the LAPW Method}}}\ (\bibinfo  {publisher} {Kluwer
  Academic Publishers, Boston},\ \bibinfo {year} {1994})\BibitemShut {NoStop}%
\bibitem [{\citenamefont {\texttt{elk.sourceforge.net}}()}]{elk}%
  \BibitemOpen
  \bibfield  {author} {\bibinfo {author} {\bibnamefont
  {\texttt{elk.sourceforge.net}}},\ }\href@noop {} {}\BibitemShut {NoStop}%
\bibitem [{\citenamefont {Perdew}\ and\ \citenamefont {Wang}(1992)}]{lda}%
  \BibitemOpen
  \bibfield  {author} {\bibinfo {author} {\bibfnamefont {J.~P.}\ \bibnamefont
  {Perdew}}\ and\ \bibinfo {author} {\bibfnamefont {Y.}~\bibnamefont {Wang}},\
  }\href@noop {} {\bibfield  {journal} {\bibinfo  {journal} {Phys. Rev. B}\
  }\textbf {\bibinfo {volume} {45}},\ \bibinfo {pages} {13244} (\bibinfo {year}
  {1992})}\BibitemShut {NoStop}%
\bibitem [{\citenamefont {Kubler}\ \emph {et~al.}(1988)\citenamefont {Kubler},
  \citenamefont {Hock}, \citenamefont {J.},\ and\ \citenamefont
  {Williams}}]{KHSW88}%
  \BibitemOpen
  \bibfield  {author} {\bibinfo {author} {\bibfnamefont {J.}~\bibnamefont
  {Kubler}}, \bibinfo {author} {\bibfnamefont {K.-H.}\ \bibnamefont {Hock}},
  \bibinfo {author} {\bibfnamefont {S.}~\bibnamefont {J.}}, \ and\ \bibinfo
  {author} {\bibfnamefont {A.~R.}\ \bibnamefont {Williams}},\ }\href@noop {}
  {\bibfield  {journal} {\bibinfo  {journal} {J. Phys. F: Met. Phys.}\ }\textbf
  {\bibinfo {volume} {18}},\ \bibinfo {pages} {469} (\bibinfo {year}
  {1988})}\BibitemShut {NoStop}%
\bibitem [{\citenamefont {Dewhurst}\ \emph {et~al.}(2016)\citenamefont
  {Dewhurst}, \citenamefont {Krieger}, \citenamefont {Sharma},\ and\
  \citenamefont {Gross}}]{Dewhurst2016}%
  \BibitemOpen
  \bibfield  {author} {\bibinfo {author} {\bibfnamefont {J.~K.}\ \bibnamefont
  {Dewhurst}}, \bibinfo {author} {\bibfnamefont {K.}~\bibnamefont {Krieger}},
  \bibinfo {author} {\bibfnamefont {S.}~\bibnamefont {Sharma}}, \ and\ \bibinfo
  {author} {\bibfnamefont {E.~K.}\ \bibnamefont {Gross}},\ }\href {\doibase
  10.1016/j.cpc.2016.09.001} {\bibfield  {journal} {\bibinfo  {journal}
  {Computer Physics Communications}\ }\textbf {\bibinfo {volume} {209}},\
  \bibinfo {pages} {92} (\bibinfo {year} {2016})}\BibitemShut {NoStop}%
\bibitem [{\citenamefont {Suzuki}\ \emph {et~al.}(2005)\citenamefont {Suzuki},
  \citenamefont {Muraoka}, \citenamefont {Inaba}, \citenamefont {Miyagawa},
  \citenamefont {Kawamura}, \citenamefont {Shimatsu}, \citenamefont {Maruyama},
  \citenamefont {Ishimatsu}, \citenamefont {Isohama},\ and\ \citenamefont
  {Sonobe}}]{Suzuki2005}%
  \BibitemOpen
  \bibfield  {author} {\bibinfo {author} {\bibfnamefont {M.}~\bibnamefont
  {Suzuki}}, \bibinfo {author} {\bibfnamefont {H.}~\bibnamefont {Muraoka}},
  \bibinfo {author} {\bibfnamefont {Y.}~\bibnamefont {Inaba}}, \bibinfo
  {author} {\bibfnamefont {H.}~\bibnamefont {Miyagawa}}, \bibinfo {author}
  {\bibfnamefont {N.}~\bibnamefont {Kawamura}}, \bibinfo {author}
  {\bibfnamefont {T.}~\bibnamefont {Shimatsu}}, \bibinfo {author}
  {\bibfnamefont {H.}~\bibnamefont {Maruyama}}, \bibinfo {author}
  {\bibfnamefont {N.}~\bibnamefont {Ishimatsu}}, \bibinfo {author}
  {\bibfnamefont {Y.}~\bibnamefont {Isohama}}, \ and\ \bibinfo {author}
  {\bibfnamefont {Y.}~\bibnamefont {Sonobe}},\ }\href {\doibase
  10.1103/PhysRevB.72.054430} {\bibfield  {journal} {\bibinfo  {journal}
  {Physical Review B - Condensed Matter and Materials Physics}\ }\textbf
  {\bibinfo {volume} {72}},\ \bibinfo {pages} {1} (\bibinfo {year}
  {2005})}\BibitemShut {NoStop}%
\end{thebibliography}
%

\end{document}